\documentclass[12pt]{article}

\makeatletter
\@ifundefined{isMain}{\def\isMain{0}}{}
\makeatother

\input{paths.tex}
\usepackage{packages}
\usepackage{commands}
\usepackage{xr-hyper}
\usepackage{hyperref}
\newcommand{\NPRtranspose}{\scriptscriptstyle T}

\makeatletter
\@ifundefined{r@Supp-NPR.sec:nonsymmetricEigendecomposition}{%
  \newlabel{Supp-NPR.sec:nonsymmetricEigendecomposition}{{S2}{2}{Real block representation}{section.2}{}}%
}{}
\@ifundefined{r@Supp-npr.sec:analyticalSolvesBlockIntegrals}{%
  \newlabel{Supp-npr.sec:analyticalSolvesBlockIntegrals}{{S3}{4}{Analytical Solves for Block Integrals}{section.3}{}}%
}{}
\@ifundefined{r@Supp-npr.sec:gradWrtAuxiliaryParameters}{%
  \newlabel{Supp-npr.sec:gradWrtAuxiliaryParameters}{{S1}{1}{\texorpdfstring {Gradient with respect to the auxiliary parameters $\NPRparameter {\NPRelementIndices }$ in partially observed CTMCs}{Gradient with respect to the auxiliary parameters in partially observed CTMCs}}{section.1}{}}%
}{}
\makeatother

\usepackage[labelfont=bf]{caption}

\begin{document}
% \linenumbers disabled for arXiv
\justifying
\captionsetup{font=doublespacing}
\doublespacing
\begin{flushright}
Version dated: \today\\
\end{flushright}

\bigskip
\medskip
\begin{center}

\noindent{\Large \bf
	Nonparametric Modeling of Continuous-Time Markov Chains
    with Scalable Exact and Approximate Gradients
}
\bigskip

\noindent{\normalsize \sc
	Filippo Monti$^{1}$ \\
  Xiang Ji$^{2}$ \\
  Yucai Shao$^{1}$ \\
  Marc A.~Suchard$^{1,3,4}$} \\

\bigskip
\noindent {\small
  \it $^1$ Department of Biostatistics, Jonathan and Karin Fielding School of Public Health, University of California Los Angeles, Los Angeles, CA, USA \\
  \it $^2$ Department of Statistics, Iowa State University, Ames, IA, USA \\
  \it $^{3}$ Department of Biomathematics, David Geffen School of Medicine at UCLA, University of California Los Angeles, Los Angeles, CA, USA \\
  \it $^{4}$ Department of Human Genetics, David Geffen School of Medicine at UCLA, University of California Los Angeles, Los Angeles, CA, USA
} \\

\end{center}
\medskip
\noindent{\bf Corresponding author:} Marc A.~Suchard, Departments of Biostatistics, Biomathematics, and Human Genetics,
University of California Los Angeles, 695 Charles E.~Young Dr., South,
Los Angeles, CA 90095-7088, USA; E-mail: \href{mailto:msuchard@ucla.edu}{msuchard@ucla.edu}

\vspace{1in}

\clearpage

\paragraph{Abstract}
Inferring the infinitesimal rates of continuous-time Markov chains (CTMCs) is a central challenge in many scientific domains.
This task is hindered by three factors: quadratic growth in the number of rates as the CTMC state space expands, strong dependencies among rates, and incomplete information for many transitions.
We introduce a new Bayesian framework that flexibly models CTMC rates by incorporating covariates through Gaussian processes (GPs).
This approach improves inference by integrating new information and deepens our understanding of CTMC stochastic behavior by shedding light on potential external drivers.
Unlike previous approaches limited to linear covariate effects, our method captures complex nonlinear relationships, enabling fuller use of covariate information and more accurate characterization of their influence.

For efficient posterior inference, we employ Hamiltonian Monte Carlo (HMC).
To avoid the prohibitive cost of repeated matrix-exponential derivatives under existing approaches, we develop two scalable strategies for exact and approximate gradient computation.
Given $\NPRndata$ observations and $\NPRnstates$ CTMC states, these strategies reduce the cost from ${\cal O}(\NPRndata \NPRnstates^3)$, with a substantial constant, to ${\cal O}(\NPRnstates^3 + \NPRndata \NPRnstates^2)$, with cheaper constants.
We demonstrate the method in Bayesian phylogenetic and phylogeographic inference, showing strong performance on synthetic and real datasets and empirical quadratic scaling in $\NPRnstates$, even when $\NPRndata < \NPRnstates$.

\vspace{1cm}

\noindent \textbf{Keywords}: Bayesian inference, Gaussian processes, Hamiltonian Monte Carlo, matrix exponential gradients, adjoint methods, phylogenetics

\clearpage

\bibliographystyle{apalike}
\setlength{\abovedisplayskip}{6pt}
\setlength{\belowdisplayskip}{6pt}
\setlength{\abovedisplayshortskip}{3pt}
\setlength{\belowdisplayshortskip}{3pt}

\section{Introduction}
Continuous-time Markov chains (CTMCs) are a probabilistic modeling tool used in a wide range of fields such as phylogenetics, epidemiology, survival analysis, and biochemistry \citep{jukes_evolution_1969, kalbfleisch_analysis_1985}.
A CTMC is a stochastic process that randomly transitions between discrete states over continuous time.
The transitions occur at exponentially distributed waiting times, and the future state depends only on the present state, not on the past (Markov property).
The parameters describing the CTMC can be collected into a matrix called an \textit{infinitesimal rate matrix}, or generator.
A key inferential task for understanding CTMC dynamics is, therefore, estimating the entries (rates) of such a matrix.
This is challenging since the number of rates increases quadratically with the number of discrete states and the data may not contain information about transitions between all pairs of states.

\paragraph{Direct parametrization} One approach to reducing the number of free parameters in the rate matrix is to impose structural restrictions.
This method has been particularly effective in fields such as evolutionary biology, where a lower-dimensional parametrization can be motivated by knowledge of the underlying biological dynamics.
A rich body of literature has explored DNA substitution processes, proposing and testing various constraints \citep{jukes_evolution_1969, kimura_simple_1980}.

Direct parametrization, however, becomes less effective when the underlying jump process is not well studied.
This issue is exacerbated as the state space grows, since, in the absence of prior knowledge, multiple parametrizations must be explored.
This, in turn, introduces a model selection problem whose complexity increases combinatorially with the size of the state space.

\paragraph{Use of covariates} Another modeling technique leverages pre-existing information captured by \textit{covariates}.
One advantage of this approach is that, as a byproduct, it sheds light on whether and how those covariates influence the CTMC jump process.
For instance, phylogeography explores the emergence and dispersal of biological entities, including pathogenic viruses, using CTMCs \citep{pybus_unifying_2012, lemey_unifying_2014}.
The rate of a viral infection spreading from one country to another is likely dependent on the distance that separates the two countries as well as their population sizes.

Within a frequentist framework, \citet{kalbfleisch_analysis_1985} proposed to study the relationship between covariates and rates based on a log-linear model.
This approach was then extended to a Bayesian setting in the context of phylogeography.
Motivated by a rich set of examples, researchers have actively contributed to the improvement of these models.
Nonetheless, the geographic locations under study tend to be numerous, and finding a credible direct parametrization that describes the movement among them remains challenging.
\paragraph{Bayesian inference} Bayesian methods have proven to be particularly well-suited for inferring CTMCs, especially when the state space is large and when simultaneous inference is essential as the CTMC is part of a larger model.
In such cases, the data may provide weak or no information for certain rate parameters.
Consequently, the use of prior distributions provides a crucial means of regularization and improved inference.

\paragraph{Sampling} There has been a flourishing literature on Markov chain Monte Carlo (MCMC) techniques \citep{metropolis_equation_1953, hastings_monte_1970} to sample from the posterior distribution of CTMC rates \citep{suchard_bayesian_2001}.
However, when the state space is large, employing a simple Metropolis-Hastings-based MCMC for sampling rates becomes challenging, not only due to their high dimensionality but also due to the strong mutual dependencies that hinder efficient exploration of the parameter space.
Adopting a Hamiltonian Monte Carlo (HMC)-based sampling scheme \citep{neal_mcmc_2011} has proven to be an effective solution for sampling high-dimensional and highly dependent parameters.
For instance, in the context of CTMCs, \citet{zhao_bayesian_2016} use HMC to sample rates based on a Bayesian log-linear model.

\paragraph{Our proposal} We introduce a Bayesian framework that flexibly models CTMC rates by incorporating covariate relationships through Gaussian processes (GPs).
This approach relaxes the restrictive log-linear form of existing models, enabling the capture of complex nonlinear relationships while preserving a direct interpretation of covariate effects on transformed rates.
Our contribution has three components.
First, as mentioned, we use GPs to define covariate-informed priors on transformed CTMC rates, providing a flexible alternative to log-linear rate models.
Second, we propose two complementary strategies, one \textit{exact} and one \textit{approximate}, for the computation of gradients of partially observed CTMC likelihoods with respect to the rate parameters, avoiding the repeated cubic operations required by state-of-the-art methods.
Third, we implement and evaluate the framework in Bayesian phylogenetic and phylogeographic inference, including synthetic and real-data examples, and show empirically through benchmarking that the proposed gradient computation strategies improve running time by an order of magnitude in $\NPRnstates$, even when $\NPRndata < \NPRnstates$.

The computational contributions are needed because HMC-based sampling for partially observed CTMC models requires repeated evaluation of the (log-) likelihood gradient with respect to the parameters of interest.
In particular,

with $\NPRnstates$ states and $\NPRndata$ observations,

this requires evaluating expensive matrix exponential derivatives at least once per observation, so existing methods scale as ${\cal O}(\NPRndata\NPRnstates^3)$ with a substantial leading constant.
This creates a severe computational bottleneck even for moderately sized state spaces.
In contrast, our exact reverse-mode gradient scales as ${\cal O}(\NPRnstates^3 + \NPRndata\NPRnstates^2)$, with a total cost of about two likelihood evaluations plus $\NPRndata\NPRnstates^2$ independent, fully parallelizable integral solves.
The approximate gradient approach, which is based on \citet{magee_random-effects_2024}, reduces the cost further to essentially two likelihood evaluations.

\paragraph{Structure of the paper} This paper is structured as follows: Sections \ref{NPR.sec:methods} and \ref{NPR.sec:Inference} present our proposed methodology, followed by an application to phylogenetic and phylogeographic inference in Section~\ref{sec:NPRPhylogeneticsApplication} and concluding discussion in Section~\ref{sec:NPRDiscussion}.
More specifically, Section~\ref{NPR.sec:methods} presents the likelihood and the prior choices.
We begin with a brief introduction to CTMCs (\ref{NPR.sec:backgroundCTMCs}), then derive the joint distribution of the observed data under different sampling scenarios, distinguishing between fully and partially observed CTMCs (\ref{NPR.sec:jointDistributionCTMCs}).
We also motivate the use of GPs as a prior on the rate matrix (\ref{NPR.sec:priorOnRateMatrixCTMCs}).
Section~\ref{NPR.sec:Inference} develops the inference framework, and motivates the use of HMC.
We also derive the formulas for the proposed exact and approximate likelihood gradient methods, and compare them with the main competitors.
Section~\ref{sec:NPRPhylogeneticsApplication} then extends the proposed framework to data with a tree-structured dependency, a setting of central importance in phylogenetics and phylogeography.
Finally, we demonstrate the performance of our method on synthetic and real datasets, and we empirically assess the scalability of the two proposed gradient methods relative to state-of-the-art competitors, including autodifferentiation-based approaches (Section~\ref{NPR.sec:results}).

\section{Methods I: Likelihood and prior}\label{NPR.sec:methods}
\subsection{Brief introduction to continuous-time Markov chains}\label{NPR.sec:backgroundCTMCs}
A (time-homogeneous) CTMC on a finite state space $\NPRstateSpace = \left\{ \NPRstate_1, \dots, \NPRstate_\NPRnstates \right\}$ ($|\NPRstateSpace|=\NPRnstates$) is a stochastic process $ \bracel \NPRctmc(\NPRtimeValue) :\NPRtimeValue \in \mathbb{R}^{+} \bracer$ taking values in $\NPRstateSpace$ and defined by the following two components:
\begin{itemize}[nosep]
    \item[1.] an \textit{initial distribution} $\NPRstationaryDistributionVector^*$ on $\NPRstateSpace$ so that for all $\NPRelementIndicesFirst \in \NPRstateSpace$, $\Prob{\NPRctmc(0) = \NPRelementIndicesFirst} = \NPRstationaryDistribution^*_\NPRelementIndicesFirst$,
    \item[2.] an \textit{infinitesimal rate matrix} $\NPRunnormedRateMatrix = (\NPRunnormedElement{\NPRelementIndices})_{\NPRelementIndicesFirst,\NPRelementIndicesSecond \in \NPRstateSpace}$ with non-negative off-diagonal elements ($\NPRunnormedElement{\NPRelementIndices} \geq 0$ for $\NPRelementIndicesFirst \neq \NPRelementIndicesSecond$), and with the diagonal elements constrained so that each row of $\NPRunnormedRateMatrix$ sums to zero, or $\NPRunnormedElement{\NPRdiagonalIndices} = -\sum_{\NPRelementIndicesSecond \neq \NPRelementIndicesFirst} \NPRunnormedElement{\NPRelementIndices}$ for every $\NPRelementIndicesFirst \in \NPRstateSpace$.
    Moreover, for all pairs of states $\NPRelementIndicesFirst,\NPRelementIndicesSecond \in \NPRstateSpace$, every $t \in \mathbb{R}^+$ such that $\Prob{\NPRctmc(\NPRtimeValue) = \NPRelementIndicesFirst} > 0$ then, for small values $h > 0$:
     \begin{align} \label{NPR.def:introTransProb}
        \Prob{\NPRctmc(\NPRtimeValue+h) = \NPRelementIndicesSecond|\NPRctmc(\NPRtimeValue) = \NPRelementIndicesFirst} = \indicatorRound{\NPRelementIndicesFirst = \NPRelementIndicesSecond} + \NPRunnormedElement{\NPRelementIndices} h + o(h)
    \end{align}
    where $\indicatorRound{\cdot}$ is the indicator function.
    Following a well-established convention, we adopt $\NPRunnormedElement{\NPRelementIndicesFirst} = -\NPRunnormedElement{\NPRdiagonalIndices} \geq 0$ to refer to the absolute value of the diagonal elements.
\end{itemize}

\noindent More intuitively, if the CTMC $\NPRctmct$ starts at some state $\NPRelementIndicesFirst$, it will leave that state after an exponentially distributed time with rate $\NPRunnormedElement{\NPRelementIndicesFirst}$ (\textit{time} component) and will jump to a different state $\NPRelementIndicesSecond$ with probability $\frac{\NPRunnormedElement{\NPRelementIndices}}{\NPRunnormedElement{\NPRelementIndicesFirst}}$ (\textit{location} component).
The exponential random variable describing the time spent in a state $\NPRelementIndicesFirst$ before leaving it has been assigned several names; in this work, we refer to it as the \textit{waiting time}, $\NPRsojournTime_{\NPRelementIndicesFirst}$.
When the CTMC is observed $\NPRndata$ times, we denote the vector of observations by $\NPRdataVector = (\NPRdata{1}, \NPRdata{2}, \dots, \NPRdata{\NPRndata})$.
We refer to the times when each value $\NPRdata{\NPRtimeIndex}$ is observed as $\NPRtimeValue_{\NPRtimeIndex}$; we collect those values in the vector $\NPRtimeVector = (\NPRtimeValue_{1}, \NPRtimeValue_{2}, \dots, \NPRtimeValue_{\NPRndata})$.
In the next section we present formulas for the joint distribution of $\NPRdataVector$ and $\NPRtimeVector$ under different observation models.

\subsection{Joint distribution of the observed data} \label{NPR.sec:jointDistributionCTMCs}
\subsubsection{Fully observed data}

The simplest scenario involves complete path observation during an interval, say $[0,\NPRtimeValue]$.
Specifically, we assume that we observe the visited states $\NPRdataVector$ and the waiting times $\mathbf{w}=(w_1, w_2, \dots, w_{\NPRndata})$, where $w_\NPRtimeIndex$ is the time spent in state $\NPRdata{\NPRtimeIndex}$ and $\sum_{\NPRtimeIndex=1}^{\NPRndata} w_{\NPRtimeIndex}= \NPRtimeValue$.
Therefore, we can write its joint distribution as:
\begin{align} \label{npr.eq:fullyObservedCtmc}
\condprob{\NPRdataVector, \mathbf{w}}{\NPRunnormedRateMatrix} = \NPRstationaryDistribution^*(\NPRdata{1})
\left[
\prod_{\NPRtimeIndex=1}^{\NPRndata-1} \NPRunnormedElement{\NPRdata{\NPRtimeIndex}}
e^{-\NPRunnormedElement{\NPRdata{\NPRtimeIndex}} w_\NPRtimeIndex}
\frac{\NPRunnormedElement{\NPRdata{\NPRtimeIndex}, \NPRdata{\NPRtimeIndex+1}}}{\NPRunnormedElement{\NPRdata{\NPRtimeIndex}}}
\right]
e^{-\NPRunnormedElement{\NPRdata{\NPRndata}} w_\NPRndata}.
\end{align}
In this formula, we are traversing the CTMC path along $[0,\NPRtimeValue]$ by multiplying the exponential densities of the waiting times, the rates of the observed jumps, and the probability of remaining in the final observed state until the end of the interval.

\subsubsection{Partially observed data}
A more realistic scenario involves a partially observed CTMC.
Specifically, the simplest case occurs when the state of the CTMC is known at time $0$ and at time $\NPRtimeValue$, while its behavior throughout the interval $(0,\NPRtimeValue)$ remains unknown.
This implies that an object of major importance is the finite-time transition probability of starting at a time $0$ in state $\NPRelementIndicesFirst$ and being in state $\NPRelementIndicesSecond$ at a later time $\NPRtimeValue\geq 0$, that is:
\begin{align}
\TransProbMatrixElement{\NPRelementIndices}{\NPRtimeValue} := \CondProb{\NPRctmc(\NPRtimeValue) = \NPRelementIndicesSecond}{\NPRctmc(0) = \NPRelementIndicesFirst}.
\end{align}
Collecting all such values in a matrix whose rows represent the starting states and columns the ending states, we obtain the \textit{finite-time transition probability matrix} $\TransProbMatrix{\NPRtimeValue}
 = \left(\TransProbMatrixElement{\NPRelementIndices}{\NPRtimeValue} \right)_{\NPRelementIndices}$.
 Taking the identity matrix $\mathbf{I}$ as the initial condition—meaning the process is certainly in its starting state at time $\NPRtimeValue=0$—one can derive a closed-form formula for $\TransProbMatrix{\NPRtimeValue}$ in terms of the infinitesimal rate matrix $\NPRunnormedRateMatrix$:
\begin{align} \label{NPR.eq:transProbUnnormed}
    \TransProbMatrix{\NPRtimeValue}
    =
    e^{\NPRtimeValue \NPRunnormedRateMatrix} := \sum_{\NPRtimeIndex=0}^\infty \frac{(\NPRtimeValue \NPRunnormedRateMatrix)^{\NPRtimeIndex}}{\NPRtimeIndex!},
\end{align}
where $e^{\NPRtimeValue \NPRunnormedRateMatrix}$ is the matrix exponential defined by its power series.

Now we consider specific cases where the data are only \textit{partially observed}.
We begin by assuming that we observe the data sequentially.
Within this framework, we distinguish when the sampling times are known and when they are not. Then, we briefly introduce the more general case where observations are not assumed to be sequential.

\paragraph{Observed sequential times} The simplest partially observed scenario arises when the sequence of observations includes both the states and the corresponding sampling times $(\NPRdataVector,\NPRtimeVector)$, but there is no guarantee that all the jumps of the CTMC are observed.
Then, we can write the joint distribution of $(\NPRdataVector,\NPRtimeVector)$ using the Markov property and the finite-time transition probabilities:
\begin{align}\label{NPR.eq:observedSequentialTimeLikelihood}
\condprob{\NPRdataVector,\NPRtimeVector}{\NPRunnormedRateMatrix} = \NPRstationaryDistribution^*(\NPRdata{1}) \prod_{\NPRtimeIndex=2}^\NPRndata \TransProbMatrixElement{\NPRdata{\NPRtimeIndex-1}, \NPRdata{\NPRtimeIndex}}{\NPRtimeValue_\NPRtimeIndex - \NPRtimeValue_{\NPRtimeIndex-1}}.
\end{align}
With the exception of specific, mostly low-dimensional cases, such as the Jukes-Cantor (JC) substitution model \citep{jukes_evolution_1969} for DNA mutations, the values $\TransProbMatrixElement{\NPRdata{\NPRtimeIndex-1}, \NPRdata{\NPRtimeIndex}}{\NPRtimeValue_\NPRtimeIndex}$ are generally not available in closed form.
However, they can be computed by evaluating the matrix exponential for the specified time interval, which requires a computational effort of cubic order (${\cal O}(\NPRnstates^3)$).

\paragraph{Unobserved sequential times} Suppose now that the realizations in $\NPRdataVector$ are sequentially observed, but observation times are \emph{unknown}.
Equation~\ref{NPR.eq:transProbUnnormed} shows that the waiting times and the infinitesimal rate matrix are jointly identifiable only up to a multiplicative factor.
We resolve this by constraining $\NPRunnormedRateMatrix$ and defining a time scale normalized by the expected number of jumps.
Let $\NPRstationaryDistributionVector = (\NPRstationaryDistribution_1, \dots, \NPRstationaryDistribution_\NPRnstates)$ be a probability mass vector over the state space $\NPRstateSpace$ representing \textit{a priori} state frequencies; $\NPRstationaryDistributionVector$ can represent $\NPRunnormedRateMatrix$'s stationary distribution,
 but need not. Let $\NPRjumpsCounter{\NPRtimeValue}{\NPRunnormedRateMatrix}$ count the (random) number of jumps during $[0, \NPRtimeValue]$ with respect to $\NPRstationaryDistributionVector$.
Its expected value is $\expectedValue{\NPRjumpsCounter{\NPRtimeValue}{\NPRunnormedRateMatrix}} = \NPRtimeValue \sum_{\NPRelementIndicesFirst \in \NPRstateSpace} \NPRstationaryDistribution_{\NPRelementIndicesFirst} \NPRunnormedElement{\NPRelementIndicesFirst}$, where $\NPRunnormedElement{\NPRelementIndicesFirst}= \sum_{\NPRelementIndicesSecond \neq \NPRelementIndicesFirst} \NPRunnormedElement{\NPRelementIndices}$.
To ensure this depends only on time, we rescale each entry of $\NPRunnormedRateMatrix$ by the normalizing constant
\begin{align}
\NPRnormConstant
= \expectedValue{\NPRjumpsCounter{1}{\NPRunnormedRateMatrix}}
= \sum_{\NPRelementIndicesFirst \in \NPRstateSpace} \NPRstationaryDistribution_{\NPRelementIndicesFirst} \NPRunnormedElement{\NPRelementIndicesFirst}
= \sum_{\NPRelementIndicesFirst \neq \NPRelementIndicesSecond} \NPRstationaryDistribution_{\NPRelementIndicesFirst} \NPRunnormedElement{\NPRelementIndices}.
\end{align}
We then denote the normalized rate matrix $\frac{1}{\NPRnormConstant}\NPRunnormedRateMatrix$ by
$\NPRnormedRateMatrix = (\NPRnormedElement{\NPRelementIndices})_{\NPRelementIndices}$.
The CTMC defined by $\NPRnormedRateMatrix$ has an expected number of jumps only dependent on time, $\expectedValue{\NPRjumpsCounter{\NPRtimeValue}{\NPRnormedRateMatrix}} = \NPRtimeValue$. $\NPRnormedRateMatrix$ and $\NPRtimeValue$ become identifiable since any attempt to rescale the rate matrix globally is neutralized by the normalization constant.
Here time is reinterpreted in terms of expected jumps: one time unit corresponds to the time necessary for one expected jump.
In phylogenetics, this is called ``evolutionary time'' versus natural time.
Assigning a prior $\prob{\NPRtimeVector}$ to the observation times, the joint distribution of the observations can be written as
\begin{align}
\condprob{\NPRdataVector}{\NPRunnormedRateMatrix} = \int_{\NPRtimeVector \in \mathbb{R}_{+}^{\NPRndata}} \condprob{\NPRdataVector, \NPRtimeVector}{\NPRunnormedRateMatrix}\prob{\NPRtimeVector} d\NPRtimeVector,
\end{align}
where $\condprob{\NPRdataVector, \NPRtimeVector}{\NPRunnormedRateMatrix}$ is computed as in Equation~\ref{NPR.eq:observedSequentialTimeLikelihood}.

\paragraph{CTMCs along directed acyclic graphs} In a more general framework, a CTMC evolves along a directed acyclic graph (DAG) $\tau = (\nodesSet, \edgesSet)$ where $\nodesSet$ is the set of nodes and $\edgesSet \subseteq \nodesSet \times \nodesSet$ is the set of edges.
We assume that there are \textit{no} colliders, that is, each node has at most one parent.
DAGs without colliders are usually referred to as \textit{polyforests}.
Connected polyforests are also called \textit{polytrees}.
The values of the CTMC are then observed on some subset of nodes.
Therefore, the likelihood $\condprob{\NPRdataVector}{\NPRunnormedRateMatrix}$ will have to be written accounting for the conditional independence structure that the graph induces.
The case where the observations are sequential falls within this framework (polytree), assuming that the nodes are located along a line.
Similarly, the case of multiple parallel sequential time series corresponds to a polyforest.
Section~\ref{sec:NPRPhylogeneticsApplication} presents a specific application in which the graph is tree-shaped and the observations lie at the tree tips.

\subsection{Prior on the infinitesimal rate matrix}\label{NPR.sec:priorOnRateMatrixCTMCs}
\paragraph{Infinitesimal rate matrix parametrization} The non-negativity constraints on the non-diagonal elements of the rate matrix $\NPRunnormedRateMatrix$, if directly modeled, prevent the use of convenient prior distributions defined on the real line and complicate the use of gradient-based samplers.
As is customary, we introduce a vector of real-valued auxiliary parameters $\NPRparameterVector$ and link them to the rates by a non-negative differentiable one-to-one transformation $g(\cdot)$, such that:
\begin{align} \label{npr.Eq:linkFunction}
\NPRunnormedElement{\NPRelementIndices} = g(\NPRparameter{\NPRelementIndices}).
\end{align}
For example, in this work, we shall examine the case where $g(\NPRparameter{}) = e^{\NPRparameter{}}$.
Furthermore, henceforth the infinitesimal rate matrix will be studied as a function of $\NPRparameterVector$, that is $\NPRunnormedRateMatrix=\NPRunnormedRateMatrix(\NPRparameterVector)$.

\paragraph{Covariate-based inference} Now, consider a single one-dimensional covariate $\mathbf{\NPRpredictor}=(\NPRpredictor_{\NPRelementIndices})_{\NPRelementIndices}$ that associates a value with each pair of states $\NPRelementIndices$.
For example, $\NPRpredictor_{\NPRelementIndices}$ could be the physical distance between two geographical states $\NPRelementIndicesFirst$ and $\NPRelementIndicesSecond$ which could affect the transition rate $\NPRunnormedElement{\NPRelementIndices}$ between them.
We can then use this covariate to inform a prior on the transformed rates $\theta_{\NPRelementIndices}$.
A standard choice is to assume a linear relationship on the transformed scale, for example $\theta_{\NPRelementIndices} = \beta \NPRpredictor_{\NPRelementIndices}$, which corresponds to a log-linear model when $g(\theta)=\exp(\theta)$.
However, this assumption can be restrictive because it forces the covariate effect to be monotone and linear on the chosen transformed scale.

\paragraph{Non-linear covariate effects}
In many applied settings, the relationship between a pairwise covariate and the transformed transition rates is unlikely to be log-linear.
A prominent example is \textit{saturation}: beyond a certain geographic distance, increasing spatial separation may have a diminishing effect on dispersal rates, as the rate already approaches zero and further distance adds little additional barrier.
Similarly, \textit{threshold effects} arise when a covariate must exceed a critical value before meaningfully influencing transitions.
For instance, in amino acid substitution models, small biochemical differences between two amino acids may have little effect on the substitution rate when the amino acids are functionally similar, whereas sufficiently large differences in charge, polarity, or volume may sharply reduce the rate of substitution \citep{zhao_bayesian_2016}.
\textit{Non-monotonic} relationships are also common: intermediate levels of a covariate (e.g.\ intermediate habitat connectivity or intermediate climatic similarity) may maximize transition rates, while both extremes suppress them.

\paragraph{Gaussian processes as a prior choice}
To relax the restrictive assumptions of the log-linear model employed in the previous literature \citep{pybus_unifying_2012, lemey_unifying_2014, zhao_bayesian_2016}, we propose to model the relationship between the parameters and the covariate using a more flexible GP prior.
This allows us to capture complex, nonlinear dependencies that a log-linear framework cannot adequately represent.

\paragraph{Background on Gaussian processes} The GP prior is a powerful Bayesian nonparametric prior that defines a distribution over more generic functions.
Beyond accommodating nonlinearities, the GP enables a principled way to model uncertainty, as it provides not only point estimates but also credible intervals for inferred relationships.
The GP is parameterized by a mean and a covariance function. The covariance function, also known as the GP \textit{kernel}, defines the dependencies between function values at different input points, controlling both the smoothness and variability of the inferred function.
By specifying an appropriate kernel, we can encode prior beliefs about the structure of the relationship, such as stationarity, periodicity, or varying degrees of smoothness.
Common choices include the squared exponential kernel, which enforces smooth variations, and the Matérn kernel, which allows more flexible roughness.

\paragraph{GP prior: one covariate} In mathematical terms we write the transformed rates $\NPRparameter{\NPRelementIndices}$ as a random function of covariates on which we assign a GP prior as follows:
\begin{align}
    \NPRparameter{\NPRelementIndices} = \NPRgpFunction(\NPRpredictor_{\NPRelementIndices})  \quad \text{with} \quad \NPRgpFunction(\cdot) \sim  \NPRgpDistribution{\boldsymbol{0}}{ \NPRgpKernel{}{\cdot}{\cdot| \NPRgpHyperparameters{}}},
\end{align}
where $\NPRgpFunction$ is distributed as a GP with zero mean and kernel $\NPRgpKernel{}{\cdot}{\cdot | \NPRgpHyperparameters{}}$ parametrized by the vector of hyperparameters $\NPRgpHyperparameters{}$.

\paragraph{GP prior: more covariates} If we consider a set of $\NPRnpred$ covariates $(\mathbf{\NPRpredictor}_1,\mathbf{\NPRpredictor}_2, \dots, \mathbf{\NPRpredictor}_\NPRnpred)$, then we can model the parameters $\NPRparameter{\NPRelementIndices}$ with an additive GP prior \citep{duvenaud_additive_2011}, that is:
\begin{align}
    \NPRparameter{\NPRelementIndices} &= \sum_{\NPRipred=1}^{\NPRnpred} \NPRgpFunction_{\NPRipred}(\NPRpredictor_{\NPRipred,\NPRelementIndices})  \quad \text{with} \quad \NPRgpFunction_\NPRipred(\cdot) \overset{\text{ind}}{\sim}  \NPRgpDistribution{\boldsymbol{0}}{ \NPRgpKernel{\NPRipred}{\cdot}{\cdot |\NPRgpHyperparameters{\NPRipred}}},
\end{align}
where all $\NPRgpFunction_{\NPRipred}$'s are independent and allowed to have different kernels $\NPRgpKernel{\NPRipred}{\cdot}{\cdot |\NPRgpHyperparameters{\NPRipred}}$.
By exploiting the independence of the $\NPRgpFunction_\NPRipred$'s, the expression can be integrated out analytically.
In vector form, this implies that $\NPRparameterVector$ follows a multivariate normal distribution,
\begin{align}
\NPRparameterVector &\sim \NPRmvnDistribution{\boldsymbol{0}}{\sum_{\NPRipred=1}^{\NPRnpred} \NPRgpKernel{\NPRipred}{\mathbf{\NPRpredictor}_\NPRipred}{\mathbf{\NPRpredictor}_\NPRipred^{\NPRtranspose}| \NPRgpHyperparameters{\NPRipred}}}.
\end{align}
This framework can then be easily generalized to account for interactions between covariates \citep{duvenaud_additive_2011}.

\section{Methods II: Inference} \label{NPR.sec:Inference}

\paragraph{Limits of traditional MCMC methods} The infinitesimal rates $\NPRunnormedElement{\NPRelementIndices}$ are governed by complex dependencies, and their number grows quadratically with the number of states $\NPRnstates$. This creates a high-dimensional parameter space with strong correlations among parameters.
As a result, traditional random-walk Markov chain Monte Carlo (MCMC) methods often mix slowly, exhibit high autocorrelation, and converge poorly.

\paragraph{Hamiltonian Monte Carlo} We adopt HMC \citep{duane_hybrid_1987}, a gradient-based sampling method that is particularly well-suited for traversing high-dimensional and strongly correlated posterior distributions.
By introducing auxiliary momentum variables and simulating Hamiltonian dynamics, HMC is able to propose distant, informed updates that preserve detailed balance while reducing random-walk behavior \citep{neal_mcmc_2011}.
This improves exploration of the parameter space, accelerates mixing, and lowers autocorrelation compared to standard MCMC methods.

\paragraph{Challenges with HMC gradient computation} Each HMC update requires the computation of the likelihood gradient.
In \textit{partially observed} CTMCs, the likelihood depends on matrix exponentials of the rate matrix.
Differentiating these matrix exponentials is computationally demanding, typically requiring at least ${\cal O}(\NPRndata\NPRnstates^3)$ operations per evaluation using state-of-the-art approaches, including autodifferentiation.
As $\NPRndata$ grows, this cost quickly becomes prohibitive even for moderate-sized CTMC state spaces.

\paragraph{Proposed solutions: novel exact and approximate gradient derivations}
To address the gradient bottleneck, we propose two complementary methods, one exact and one approximate, that achieve an overall complexity of $\mathcal{O}(\NPRnstates^3 + \NPRndata\NPRnstates^2)$ by exploiting the eigenstructure of the rate matrix.
The \textit{approximate} method adopts a scalable matrix exponential derivative approximation \citep{magee_random-effects_2024}, which reduces the per-edge gradient cost from $\mathcal{O}(\NPRnstates^3)$ to $\mathcal{O}(\NPRnstates^2)$ for all rate parameters simultaneously.

This approximation is embedded in an HMC scheme with surrogate trajectories \citep{rasmussen_gaussian_2003, fielding_efficient_2011, li_neural_2019}, where a Metropolis--Hastings correction step guarantees that the exact posterior remains the stationary distribution.

The \textit{exact} method is instead based on a new analytical closed form for reverse-mode likelihood gradients via the adjoint Fréchet derivative of the matrix exponential.
It matches the approximate method's total complexity, at the additional cost of $\NPRnstates^2$ block integral solves per edge.
These solves are fully parallelizable both within and between edges.

The next sections are organized as follows.
In Section~\ref{NPR.sec:likelihoodEvaluation}, we first describe a representation of the CTMC likelihood and show that, under a block-diagonal eigendecomposition of the rate matrix $\NPRnormedRateMatrix$, it can be evaluated in $\mathcal{O}(\NPRnstates^3 + \NPRndata \NPRnstates^2)$ operations.
This construction serves as the foundation for the gradient derivations.
We then review state-of-the-art approaches for computing matrix exponential gradients (Section~\ref{npr.sec:exactGradStateofTheArt}), before introducing the proposed approximate (Section~\ref{npr.sec:approxGrad}) and exact adjoint-based method (Section~\ref{NPR.sec:exactAdjointGradient}).

\subsection{Likelihood evaluation} \label{NPR.sec:likelihoodEvaluation}
Consider a rate matrix $\NPRnormedRateMatrix$ and assume that it is non-defective, equivalently, diagonalizable over $\mathbb{C}$.
This assumption is generic in the sense that non-defective square matrices form a dense subset of the space of square matrices, whereas defective matrices constitute a set of Lebesgue measure zero \citep{meyer_matrix_2023}.
Since $\NPRnormedRateMatrix$ is real, its complex conjugate eigenpairs can be represented by real two-dimensional invariant subspaces \citep{golub_matrix_2013}.
Thus, $\NPRnormedRateMatrix$ admits a block-diagonal representation of the form
\begin{align} \label{npr.eq:blockdiagonalrepresentation}
    \NPRnormedRateMatrix =  \NPRRotation  \NPRBlkDiag  \NPRRotation^{-1},
\end{align}
where $\NPRBlkDiag$ is a block diagonal matrix with $1\times1$ blocks for real eigenvalues and $2\times2$ blocks for complex conjugate eigenvalues.
The matrix $\NPRRotation$ is a real change-of-basis matrix whose columns form a real invariant-subspace basis: columns associated with real eigenvalues are real eigenvectors, while each pair of columns associated with a complex conjugate eigenpair spans the corresponding two-dimensional real invariant subspace (see Supplementary Section~\ref{Supp-NPR.sec:nonsymmetricEigendecomposition} for more details).
The matrix exponential associated with a time $\NPRtimeValue$ can then be written as
\begin{align} \label{NPR.eq:blockDiagonalReprentationOfMatrixExponential}
    e^{\NPRtimeValue \NPRnormedRateMatrix} =  \NPRRotation  e^{\NPRtimeValue\NPRBlkDiag } \NPRRotation^{-1}.
\end{align}
\noindent This spectral representation is the computational foundation of both proposed gradient-computation strategies.

The likelihood of a partially observed CTMC on a DAG can be built from edge-wise matrix--vector products.
Each product combines a finite-time transition matrix with a vector representing a possibly unnormalized distribution over states.
In particular, consider an edge of length $\NPRtimeValue$ and let $\NPRprobabilityVector$ denote a distribution vector associated with one of the nodes at the extremity of the edge.
If the state of the node is observed, then $\NPRprobabilityVector$ is a Dirac measure concentrated on the observed state.
Otherwise, $\NPRprobabilityVector$ represents either a conditional distribution or a likelihood vector summarizing the uncertainty about the latent state at that node given the observations in the corresponding subgraph.
Assume that, for example, $\NPRprobabilityVector$ is the vector associated with the child of the edge.
Then, the matrix--vector product that integrates uncertainty along the edge has the form
\begin{align}\label{NPR.eq:matriVectorProduce}
    \TransProbMatrix{\NPRtimeValue}\NPRprobabilityVector = \NPRRotation  e^{\NPRtimeValue\NPRBlkDiag } \NPRRotation^{-1} \NPRprobabilityVector,
\end{align}
which can be evaluated in quadratic time $\mathcal{O}(\NPRnstates^2)$ by following the order of multiplications from right to left.

In the general framework where the CTMC evolves along a DAG with no colliders (each node has at most one parent), evaluating the likelihood
requires computing operation~\eqref{NPR.eq:matriVectorProduce} once for each edge.
Assuming that every internal node has at least two children, then the number of edges is linear in the number of observations $\NPRndata$.
Therefore, the full likelihood can be evaluated in
\begin{align}
    \mathcal{O}(\NPRnstates^3 + \NPRndata \NPRnstates^2)
\end{align}
time, where the cubic cost is paid only once for the eigendecomposition.

\paragraph{Likelihood representation for polytrees (connected DAGs with no colliders)} We now consider a DAG with no colliders that is  \textit{connected}, meaning that any two nodes are joined by a path when edge directions are ignored.
Let $\NPRchildNode$ be an internal node, $\text{pa}(\NPRchildNode)$ be its parent, and $\NPRtimeValue_{\NPRchildNode}$ be the length of the edge connecting them.
For each node $\NPRchildNode$, denote by $\NPRdata{\lceil \NPRchildNode \rceil}$ the set of observations lying in the upper graph induced by $\NPRchildNode$, that is, its non-descendants.
Similarly, let $\NPRdata{\lfloor \NPRchildNode \rfloor}$ be the set of descendant observations of $\NPRchildNode$.
Without loss of generality, as we will explain in the next paragraph, assume that the data $\NPRdataVector = (\NPRdata{1}, \NPRdata{2}, \dots, \NPRdata{\NPRndata})$ are observed at nodes with no children.
Then, their joint distribution can be expressed by conditioning on the latent states $(\NPRdata{\NPRchildNode},\NPRdata{\text{pa}(\NPRchildNode)})$:
\begin{align} \label{NPR.eq:characterizationTDLPrePostOrder}
    \prob{\NPRdata{1}, \dots, \NPRdata{\NPRndata}}
    &= \expectedValue{\condprob{\NPRdata{1}, \dots, \NPRdata{\NPRndata}}{\NPRdata{\text{pa}(\NPRchildNode)}, \NPRdata{\NPRchildNode}} } \quad  \text{by the law of iterated expectations} \nonumber \\
    &= \expectedValue{\condprob{\NPRdata{\lceil \NPRchildNode \rceil}}{\NPRdata{\text{pa}(\NPRchildNode)}} \condprob{\NPRdata{\lfloor \NPRchildNode \rfloor}}{\NPRdata{\NPRchildNode}}}
    \quad \text{by conditional independence} \nonumber \\
    &= \sum_{\NPRdata{\text{pa}(\NPRchildNode)} \in \NPRstateSpace} \sum_{\NPRdata{\NPRchildNode} \in \NPRstateSpace}
    \left[ \condprob{\NPRdata{\lceil \NPRchildNode \rceil}}{\NPRdata{\text{pa}(\NPRchildNode)}} \condprob{\NPRdata{\lfloor \NPRchildNode \rfloor}}{\NPRdata{\NPRchildNode}} \right] \condprob{\NPRdata{\NPRchildNode}}{\NPRdata{\text{pa}(\NPRchildNode)}} \prob{\NPRdata{\text{pa}(\NPRchildNode)}} & \nonumber \\
    &= \sum_{\NPRdata{\text{pa}(\NPRchildNode)} \in \NPRstateSpace} \sum_{\NPRdata{\NPRchildNode} \in \NPRstateSpace} \prob{\NPRdata{\lceil \NPRchildNode \rceil}, \NPRdata{\text{pa}(\NPRchildNode)}} \condprob{\NPRdata{\NPRchildNode}}{\NPRdata{\text{pa}(\NPRchildNode)}} \condprob{\NPRdata{\lfloor \NPRchildNode \rfloor}}{\NPRdata{\NPRchildNode}} & \nonumber \\
    &= \NPRpreOrder{\NPRchildNode}^{\NPRtranspose} \, \TransProbMatrix{\NPRbranchLength{\NPRchildNode}} \NPRpostOrder{\NPRchildNode}^{\vphantom{\NPRtranspose}},
\end{align}
where, in the last equality, we note that $\condprob{\NPRdata{\NPRchildNode}}{\NPRdata{\text{pa}(\NPRchildNode)}} = \TransProbMatrixElement{\NPRdata{\text{pa}(\NPRchildNode)}\NPRdata{\NPRchildNode}}{\NPRbranchLength{\NPRchildNode}}$, and
\begin{itemize}
    \item $\NPRpostOrder{\NPRchildNode}$ is the \textit{post-order partial likelihood} vector evaluated at the child node $\NPRchildNode$ containing $\condprob{\NPRdata{\lfloor \NPRchildNode \rfloor}}{\NPRdata{\NPRchildNode} = \NPRstate_k}$ for each state $\NPRstate_k \in \NPRstateSpace$, and
    \item $\NPRpreOrder{\NPRchildNode}$ is the \textit{pre-order partial likelihood} vector evaluated at the parent node $\text{pa}(\NPRchildNode)$ containing $\prob{\NPRdata{\lceil \NPRchildNode \rceil}, \NPRdata{\text{pa}(\NPRchildNode)} = \NPRstate_k}$ for each state.
    Notice that $\NPRpreOrder{\NPRchildNode}$ is edge-indexed: it denotes the pre-order vector at the parent end of the edge leading to $\NPRchildNode$.
\end{itemize}

When observations also occur at internal nodes, that is, at nodes with children, they introduce independence between the upper and lower graphs.
This implies that the joint distribution can be factorized, and each subgraph can be represented separately by \eqref{NPR.eq:characterizationTDLPrePostOrder}.
For example, when observations are sequential, the joint likelihood can be written as the product
\begin{align}
    \prod_{\NPRchildNode=2}^{\NPRndata}
    \NPRpreOrder{\NPRchildNode}^{\NPRtranspose} \, \TransProbMatrix{\NPRbranchLength{\NPRchildNode}} \NPRpostOrder{\NPRchildNode}^{\vphantom{\NPRtranspose}},
\end{align}
where both $\NPRpreOrder{\NPRchildNode}$ and $\NPRpostOrder{\NPRchildNode}$ correspond to observations at the beginning and end of the edge, that is, they are Dirac measures on the observed states.

\paragraph{Message-passing algorithm for polytrees}
When the graph includes unobserved nodes, we need to compute $\NPRpostOrder{\NPRchildNode}$ and $\NPRpreOrder{\NPRchildNode}$ iteratively by passing information from observed nodes along the graph \citep{felsenstein_evolutionary_1981}.
In particular, the post-order vector at a node $\NPRchildNode$ that has children in the set $\nodesSet_{\text{child}(\NPRchildNode)} \subset \nodesSet$ can be built using the following expression:
\begin{equation}
\NPRpostOrder{\NPRchildNode} = \!\! \!\!  \!\! \!\!\mathop{\otimes}\limits_{\quad \enspace \NPRparentNode \in \nodesSet_{\text{child}(\NPRchildNode)}}   \!\!\!\!\!\!\TransProbMatrix{\NPRbranchLength{\NPRparentNode} } \NPRpostOrder{\NPRparentNode}
\label{NPR.eq:pruningPostOrder}
\end{equation}
where $\otimes$ denotes element-wise multiplication, and $\NPRbranchLength{\NPRparentNode}$ is the length of the edge connecting $\NPRparentNode$ to its parent $\text{pa}(\NPRparentNode) = \NPRchildNode$.
The intuition is straightforward: we propagate the distribution vectors up each edge using the corresponding finite-time transition matrices, and then combine the resulting vectors through element-wise multiplication.

Similarly, the algorithm for the pre-order partial likelihoods runs from the root to the tips by applying:
\begin{equation}
\NPRpreOrder{\NPRchildNode} =
\TransProbMatrix{\NPRbranchLength{\text{pa}(\NPRchildNode)}}^{\NPRtranspose} \NPRpreOrder{\text{pa}(\NPRchildNode)} \Hadamard \left[ \!\! \!\!\mathop{\otimes}\limits_{\quad \; \NPRparentNode \in \nodesSet_{\text{sibl}(\NPRchildNode)}} \!\! \TransProbMatrix{\NPRbranchLength{\NPRparentNode}}\NPRpostOrder{\NPRparentNode}\right].
\label{NPR.eq:pruningPreOrder}
\end{equation}
where $\nodesSet_{\text{sibl}(\NPRchildNode)} \subset \nodesSet$ is the set of siblings of $\NPRchildNode$ and $\Hadamard$ is the element-wise product.
This formula allows to compute the conditional likelihood of the upper graph by merging the information coming from the siblings and from the parent.
The algorithm is initialized by setting $\NPRpreOrder{\text{root}}$ to the chosen root-state distribution, for example, the stationary distribution of the CTMC.

\paragraph{Noisy observations} Noisy observations are the norm in many applied settings.
The likelihood-vector representation above accommodates noisy or probabilistic observations without affecting the asymptotic computational cost.
A formal way to model noise is to add an edge between the latent value and the observed one, and assume the observation is generated by transforming the latent vector with some (time-independent) stochastic matrix acting along the edge.
For example, if we want to add a small perturbation, we can multiply the probability vectors by a stochastic matrix with most of the probability mass concentrated on the diagonal.
This matrix can also contain unknown random elements and be observation-specific.

\subsection{Likelihood gradient computations}

The difficulty of computing the likelihood gradient with respect to the transformed (unnormalized) rates $\NPRparameter{\NPRelementIndices} = g^{-1}(\NPRunnormedElement{\NPRelementIndices})$ varies depending on the sampling scenario.
For fully observed CTMCs, the gradient is straightforward from Equation~\eqref{npr.eq:fullyObservedCtmc}, so it is omitted here.
In contrast, for partially observed CTMCs, $\NPRmyFunc$ involves multiple matrix exponentials that must be differentiated, as shown by Equation~\eqref{NPR.eq:characterizationTDLPrePostOrder}.
We first review why standard forward- and reverse-mode approaches require repeated cubic operations, leading to at least $\mathcal{O}(\NPRndata\NPRnstates^3)$ operations with substantial per-edge leading constant factors.
We then develop two alternatives: an approximate forward-mode method and an exact reverse-mode method.
Both exploit the block-diagonal eigendecomposition of the rate matrix in Equation~\eqref{npr.eq:blockdiagonalrepresentation} and reduce the computational complexity to only $\mathcal{O}(\NPRnstates^3 + \NPRndata\NPRnstates^2)$.
Section~\ref{Supp-npr.sec:gradWrtAuxiliaryParameters} of the Supplement presents the specific derivations of the gradient with respect to the parameters $\NPRparameter{\NPRelementIndices}$ in the presence of rate-dependent normalization constants.

\subsubsection{Forward- and reverse-mode gradients of the matrix exponential: state of the art} \label{npr.sec:exactGradStateofTheArt}
\citet{najfeld_derivatives_1995} review several strategies to represent and compute the Fréchet derivative $\NPRFrechetDerivative{\NPRnormedRateMatrix}{\NPRFrechetDirection}$ of the matrix exponential $e^{\NPRtimeValue \NPRnormedRateMatrix}$ along a generic direction $\NPRFrechetDirection$.
We review three of them and refer the reader to that work for the others:

\begin{itemize}
    \item[1.] \textit{Block-matrix augmentation representation} \citep{vanloan_computing_1978a}:
    one of the most popular and straightforward approaches relies on building the block matrix
    \begin{align} \label{npr.eq:exactGradientKal}
    \begin{bmatrix}
        \NPRnormedRateMatrix & \NPRFrechetDirection \\
        \mathbf{0} & \NPRnormedRateMatrix
    \end{bmatrix},
\end{align}
scaling it by $\NPRtimeValue$ and exponentiating it. The Fréchet derivative along a direction $\NPRFrechetDirection$ corresponds to the upper-right block of the resulting matrix.

\item[2.] \textit{Series representation}:
\begin{align} \label{NPR.eq:commutatorDefinitionGradientMatrixExponential}
    \NPRFrechetDerivative{\NPRnormedRateMatrix}{\NPRFrechetDirection} =  e^{t\NPRnormedRateMatrix} \sum_{k=0}^\infty \frac{t^{k+1}}{(k+1)!} \left\{\NPRFrechetDirection, \NPRnormedRateMatrix^k \right\},
\end{align}
where $\left\{\NPRFrechetDirection, \NPRnormedRateMatrix^k \right\}$ is defined recursively as
$\left\{\NPRFrechetDirection, \NPRnormedRateMatrix^0 \right\}=\NPRFrechetDirection$ and $\left\{\NPRFrechetDirection, \NPRnormedRateMatrix^{k} \right\} = \Big[\left\{\NPRFrechetDirection, \NPRnormedRateMatrix^{k-1} \right\}, \NPRnormedRateMatrix \Big]$,
where $[\cdot, \cdot]$ is the matrix commutator.

\item[3.] \textit{Integral representation}:
\begin{align}\label{NPR.eq:integralRepresentationFrechetDerivative}
\NPRFrechetDerivative{\NPRnormedRateMatrix}{\NPRFrechetDirection} = \NPRtimeValue
\int_0^1
e^{(1-s)\NPRtimeValue\NPRnormedRateMatrix}
\NPRFrechetDirection\,
e^{s\NPRtimeValue\NPRnormedRateMatrix}
\dd s.
\end{align}
\end{itemize}

\noindent Since there is a scalar output (the log-likelihood) and about $\NPRnstates^2$ parameters, reverse-mode differentiation is the more natural approach \citep{baydin_automatic_2018}.
Nonetheless, we first illustrate the familiar forward-mode formulation before turning to the more appropriate reverse-mode counterpart.
\paragraph{Forward-mode gradients}
To compute the forward-mode gradient, we need to evaluate the Fréchet derivatives in the direction of each element $(\NPRelementIndicesFirst, \NPRelementIndicesSecond)$ of the matrix $\NPRnormedRateMatrix$.
In particular, for each $(\NPRelementIndicesFirst, \NPRelementIndicesSecond)$ we compute $\NPRFrechetDerivative{\NPRnormedRateMatrix}{\NPRFrechetDirection_{\NPRelementIndices}}$, where $\NPRFrechetDirection_{\NPRelementIndices}$ has the $(\NPRelementIndicesFirst, \NPRelementIndicesSecond)$-th element equal to one and all others set to zero.
Since the cost of each operation is cubic and $\NPRnormedRateMatrix$ has $\NPRnstates^2$ entries, the full exact forward gradient costs $\mathcal{O}(\NPRnstates^5)$.
This operation has to be repeated at every edge, leading to a total cost scaling of $\mathcal{O}(\NPRndata \NPRnstates^5)$.

\paragraph{Reverse-mode gradients}
Reverse-mode differentiation replaces the $\NPRnstates^2$ repeated forward passes with only one forward and one backward pass, reducing the time complexity to $\mathcal{O}(\NPRndata \NPRnstates^3)$.
A general reverse-mode approach would proceed as follows.
First, compute the adjoint of the finite-time transition matrices at each edge.
Then, backpropagate through the matrix exponential to obtain edge-specific rate-matrix adjoints, requiring $\mathcal{O}(\NPRndata\NPRnstates^3)$ operations.
Finally, accumulate the contributions across edges.

\newcommand{\NPRDiagMatrix}{\mathbf{D}}
\newcommand{\NPRDiagMatrixElement}[1]{d_{#1}}
\paragraph{Reverse-mode gradients under reversibility} \citet{lieser_phylograd_2025} recently implemented an analytical derivation of reverse-mode gradients for phylogenetic likelihoods under reversibility.
Reversibility ensures that the rate matrix $\NPRnormedRateMatrix$ is similar to a symmetric matrix.
This yields a real diagonalization $\NPRnormedRateMatrix = \NPRRotation \NPRDiagMatrix \NPRRotation^{-1}$, where $\NPRDiagMatrix$ is diagonal and $\NPRRotation$ is invertible.
With this diagonalization, one can use the following closed-form divided-difference formula for the Fréchet derivative \citep{najfeld_derivatives_1995}:
\begin{align}
    \NPRFrechetDerivative{\NPRnormedRateMatrix}{\NPRFrechetDirection}=
    \NPRRotation
    \left(\NPRRotation^{-1}\NPRFrechetDirection\NPRRotation
    \Hadamard
    \mathbf{X}(\NPRDiagMatrix, \NPRtimeValue)\right)
    \NPRRotation^{-1},
\end{align}
where $\Hadamard$ is the element-wise product, and $\mathbf{X}(\NPRDiagMatrix, \NPRtimeValue)$ is the symmetric divided-differences matrix with diagonal elements $\NPRtimeValue
e^{
\NPRtimeValue
\NPRDiagMatrixElement{\NPRelementIndicesFirst}
}$
and off-diagonal elements
$\dfrac{
e^{\NPRtimeValue
\NPRDiagMatrixElement{\NPRelementIndicesFirst}}
-
e^{\NPRtimeValue
\NPRDiagMatrixElement{\NPRelementIndicesSecond}}
}{\NPRDiagMatrixElement{\NPRelementIndicesFirst}
- \NPRDiagMatrixElement{\NPRelementIndicesSecond}
}$.
Denoting by $\NPRadjointVar$ the edge-specific adjoint of the finite-time transition matrix, the adjoint Fréchet derivative is equal to $[\NPRFrechetDerivative{\NPRnormedRateMatrix}{\NPRadjointVar^{\NPRtranspose}}]^{\NPRtranspose}$.
Due to the matrix multiplications, this approach remains $\mathcal{O}(\NPRndata\NPRnstates^3)$, but substantially reduces the constant factor by avoiding generic backpropagation through the matrix exponentials.

\paragraph{Automatic differentiation}
Automatic differentiation provides a general-purpose way to obtain derivatives of an implemented likelihood algorithm by recursively applying the chain rule to the elementary operations in the computation \citep{baydin_automatic_2018}.
In the present setting, one may implement the message-passing algorithm together with the edge-specific matrix exponentials, and then use reverse-mode automatic differentiation to propagate adjoints from the scalar log-likelihood back to the parameters of $\NPRnormedRateMatrix$.
This approach is attractive because it requires little model-specific algebra and can accommodate flexible parametrizations of the rate matrix.
Automatic differentiation has recently been used for matrix-exponential gradients in phylogenetic inference, for example, in \texttt{treeflow} \citep{swanepoel_treeflow_2022} and \texttt{torchtree} \citep{fourment_torchtree_2026}.
Nevertheless, \citet{fourment_automatic_2023} discuss how generic automatic differentiation tends to underperform analytical methods when the likelihood has exploitable structure, as in phylogenetic models with tree-structured conditional independence.
Similarly, in the reversible setting, \citet{lieser_phylograd_2025} show that their specialized analytical reverse-mode implementation can be substantially faster than a generic automatic-differentiation implementation.

\paragraph{Conditional gradients}
\citet{zhao_bayesian_2016} provide a related data-augmentation approach.
Rather than differentiating the marginal data likelihood directly, they introduce auxiliary CTMC trajectories.
Specifically, endpoint-conditioned trajectories between observations are sampled using Poisson-uniformization-based substitution mapping \citep{hobolth_simulation_2009} and then summarized by sufficient statistics, namely transition counts and dwell times.
Conditional on these statistics, gradients with respect to the rate-matrix parameters are inexpensive to compute because the likelihood has the fully observed form in Equation~\eqref{npr.eq:fullyObservedCtmc}.
Thus, the approach shifts the computational burden from differentiating the marginal partially observed likelihood to simulating unobserved states and computing the sufficient statistics.
While this yields fast conditional gradient evaluations, each HMC trajectory is defined with respect to an augmented conditional likelihood.
The resulting auxiliary-variable kernel still targets the marginal posterior after trajectories are refreshed and discarded, but marginal mixing may be slowed when the imputed substitution histories are strongly coupled to the rate parameters \citep{liu_covariance_1994, papaspiliopoulos_general_2007}.

\subsubsection{Gradient of the matrix exponential: Scalable forward-mode approximation}\label{npr.sec:approxGrad}

Implementing Equation~\eqref{NPR.eq:commutatorDefinitionGradientMatrixExponential} requires truncating the infinite series;
\citet{didier_surprising_2024} demonstrate that retaining only the first term of the series already provides a remarkably accurate first-order approximation:
\begin{align} \label{NPR.eq:approximationDerivativeExponentialMatrix}
    \myPartial{e^{\NPRtimeValue \NPRnormedRateMatrix}}{\NPRnormedElement{\NPRelementIndices}} \approx \NPRtimeValue e^{\NPRtimeValue \NPRnormedRateMatrix} \NPRFrechetDirection_{\NPRelementIndices}.
\end{align}
The approximation is expected to be most accurate when the omitted higher-order commutator terms in Equation~\eqref{NPR.eq:commutatorDefinitionGradientMatrixExponential} are small, for example when edge lengths are short on the normalized time scale or when the perturbation direction nearly commutes with the rate matrix.
Following surrogate-gradient HMC schemes \citep{rasmussen_gaussian_2003, fielding_efficient_2011, li_neural_2019}, the leapfrog trajectory is generated with the approximate gradient, but the final Metropolis--Hastings accept--reject probability is computed using the exact log posterior, ensuring this remains the target stationary distribution.

Now let $\NPRmyFunc$ be the CTMC likelihood on a DAG that we want to differentiate with respect to $\NPRnormedElement{\NPRelementIndices}$.
By a simple application of the chain rule, we can differentiate $\NPRmyFunc$ by summing up the effect of a change in $\NPRnormedElement{\NPRelementIndices}$ along each edge.
This should become particularly intuitive by considering the likelihood representation in Equation~\eqref{NPR.eq:characterizationTDLPrePostOrder}.
We can therefore write the derivative as follows:
\begin{align}  \label{NPR.eq:approximateForwardCTMCGradientDerivation}
    \myPartial{\NPRmyFunc}{\NPRnormedElement{\NPRelementIndices}}
    &= \sum_{\NPRchildNode}\NPRpreOrder{\NPRchildNode}^{\NPRtranspose}
     \left[\myPartial{}{\NPRnormedElement{\NPRelementIndices}} \TransProbMatrix{\NPRbranchLength{\NPRchildNode}} \right]
     \NPRpostOrder{\NPRchildNode} \nonumber\\
     &= \sum_{\NPRchildNode}\NPRpreOrder{\NPRchildNode}^{\NPRtranspose}
     \left[\myPartial{}{\NPRnormedElement{\NPRelementIndices}} e^{\NPRbranchLength{\NPRchildNode}\NPRnormedRateMatrix} \right]
     \NPRpostOrder{\NPRchildNode} \nonumber\\
     & \approx \sum_{\NPRchildNode}\NPRpreOrder{\NPRchildNode}^{\NPRtranspose}
     \left[\NPRbranchLength{\NPRchildNode} e^{\NPRbranchLength{\NPRchildNode}\NPRnormedRateMatrix} \NPRFrechetDirection_{\NPRelementIndices}\right]
     \NPRpostOrder{\NPRchildNode} \nonumber \\
     &= \sum_{\NPRchildNode}\NPRbranchLength{\NPRchildNode} \left[ e^{\NPRbranchLength{\NPRchildNode}\NPRnormedRateMatrix^{\NPRtranspose} } \NPRpreOrder{\NPRchildNode}
      \right]^{\NPRtranspose}
      \NPRFrechetDirection_{\NPRelementIndices}
     \NPRpostOrder{\NPRchildNode}
\end{align}
where, in the third line, we have plugged in the approximation.
If we adopt the block diagonal representation of the matrix exponential introduced in Equation~\eqref{NPR.eq:blockDiagonalReprentationOfMatrixExponential}, the quantities $\NPRpreOrderTemporary{\NPRchildNode} := e^{\NPRbranchLength{\NPRchildNode}\NPRnormedRateMatrix^{\NPRtranspose} } \NPRpreOrder{\NPRchildNode}$ and
$\NPRpostOrder{\NPRchildNode}$
can be computed across all edges in $\mathcal{O}(\NPRnstates^3 + \NPRndata\NPRnstates^2)$.
Then, with a simple rearrangement of Equation~\eqref{NPR.eq:approximateForwardCTMCGradientDerivation} for every pair $(\NPRelementIndicesFirst, \NPRelementIndicesSecond)$, we can write the likelihood's gradient with respect to $\NPRnormedRateMatrix$ as the following sum of outer products weighted by the edge lengths:
\begin{align}
     \myPartial{\NPRmyFunc}{\NPRnormedRateMatrix} = \sum_{\NPRchildNode} \NPRbranchLength{\NPRchildNode}
     \NPRpreOrderTemporary{\NPRchildNode} \NPRpostOrder{\NPRchildNode}^{\NPRtranspose}.
\end{align}
Notice that, if the pre-order at the root $\NPRstationaryDistributionVector_{\text{root}}$ is computed as a function of the rate matrix, then we also need to add its derivative.
A closed-form expression in the case where $\NPRstationaryDistributionVector_{\text{root}}$  is the stationary distribution is provided in \citet{kalbfleisch_analysis_1985}.
Since the accumulation of the outer product is dominated by the post- and pre-order algorithms necessary to compute and cache $\NPRpreOrderTemporary{\NPRchildNode}$ and
$\NPRpostOrder{\NPRchildNode}$, the total cost per gradient evaluation corresponds to about two likelihood evaluations, that is, $\mathcal{O}(\NPRnstates^3 + \NPRndata\NPRnstates^2)$.
The first two columns of Table~\ref{NPR.tab:complexity} summarize these costs and caching steps.

\begin{table}[ht]
\centering
\begin{tabular}{l cc cc}
\toprule
& \multicolumn{2}{c}{\textbf{Approximate}}
& \multicolumn{2}{c}{\textbf{Exact}} \\
\cmidrule(lr){2-3} \cmidrule(lr){4-5}
& $\mathcal{O}(\cdot)$ & Caching
& $\mathcal{O}(\cdot)$ & Caching \\
\midrule
One Eigendecomp.
  & $\NPRnstates^3$
  & $\NPRRotation, \NPRBlkDiag, \NPRRotation^{-1}$
  & $\NPRnstates^3$
  & $\NPRRotation, \NPRBlkDiag, \NPRRotation^{-1}$ \\[4pt]
Postorder
  & $\NPRndata\NPRnstates^2$
  & $\NPRpostOrder{\NPRchildNode}$
  & $\NPRndata\NPRnstates^2$
  & $\NPRRotation^{-1} \NPRpostOrder{\NPRchildNode} \quad $ \\[4pt]
Preorder
  & $\NPRndata\NPRnstates^2$
  & $e^{\NPRbranchLength{\NPRchildNode}\NPRnormedRateMatrix^{\NPRtranspose} } \NPRpreOrder{\NPRchildNode}$
  & $\NPRndata\NPRnstates^2$
  & $\NPRRotation^{\NPRtranspose} \NPRpreOrder{\NPRchildNode}$ \\[4pt]
Integral solves
  &
  &
  & $\NPRndata\NPRnstates^2$
  &  \\[4pt]
Outer products
  &  $\NPRndata\NPRnstates^2$
  &
  & $\NPRndata\NPRnstates^2$
  &  \\
\bottomrule
\end{tabular}
\caption{\textbf{Time complexity and caching for approximate and exact methods.}
After a one-time eigendecomposition, the \textit{approximate} method has a total computational cost comparable to two likelihood evaluations.
The \textit{exact} method additionally requires $\NPRndata\NPRnstates^2$ block integral solves, although these operations are fully parallelizable both across edges and within individual edge computations.}
\label{NPR.tab:complexity}
\end{table}

\subsubsection{A new scalable exact analytic derivation for reverse-mode gradients} \label{NPR.sec:exactAdjointGradient}

Let $\NPRFrechetDerivative{\NPRnormedRateMatrix}{\cdot} $ denote the Fr\'echet derivative of the matrix exponential.
The adjoint Fr\'echet derivative $\NPRFrechetAdj{\NPRnormedRateMatrix}{\cdot}$ is defined implicitly with respect to the Frobenius inner product by the identity
\begin{align}
\left\langle
\NPRFrechetDerivative{\NPRnormedRateMatrix}{\NPRFrechetDirection},
\NPRadjointVar
\right\rangle
=
\left\langle
\NPRFrechetDirection,
\NPRFrechetAdj{\NPRnormedRateMatrix}{\NPRadjointVar}
\right\rangle,
\end{align}
for all matrices $\NPRFrechetDirection$ and $\NPRadjointVar$.
Using the integral representation of the Fr\'echet derivative~\eqref{NPR.eq:integralRepresentationFrechetDerivative}, the definition of Frobenius inner product, and the cyclic property of the trace operator, we can see that:
\begin{align}
\left\langle
\NPRFrechetDerivative{\NPRnormedRateMatrix}{\NPRFrechetDirection},
\NPRadjointVar
\right\rangle
&=
\tr\!\left(
\NPRFrechetDerivative{\NPRnormedRateMatrix}{\NPRFrechetDirection}^{\NPRtranspose}
\NPRadjointVar
\right) \nonumber\\
&=
\NPRtimeValue
\int_0^1
\tr\!\left(
e^{s\NPRtimeValue\NPRnormedRateMatrix^{\NPRtranspose}}
\NPRFrechetDirection^{\NPRtranspose}
e^{(1-s)\NPRtimeValue\NPRnormedRateMatrix^{\NPRtranspose}}
\NPRadjointVar
\right)
\dd s \nonumber\\
&=
\NPRtimeValue
\int_0^1
\tr\!\left(
\NPRFrechetDirection^{\NPRtranspose}
e^{(1-s)\NPRtimeValue\NPRnormedRateMatrix^{\NPRtranspose}}
\NPRadjointVar
e^{s\NPRtimeValue\NPRnormedRateMatrix^{\NPRtranspose}}
\right)
\dd s \nonumber\\
&=
\tr\!\left(
\NPRFrechetDirection^{\NPRtranspose}
\Biggl[
\NPRtimeValue
\int_0^1
e^{(1-s)\NPRtimeValue\NPRnormedRateMatrix^{\NPRtranspose}}
\NPRadjointVar
e^{s\NPRtimeValue\NPRnormedRateMatrix^{\NPRtranspose}}
\dd s
\Biggr]
\right) \nonumber\\
&= \left\langle
\NPRFrechetDirection,
\NPRFrechetDerivative{\NPRnormedRateMatrix^{\NPRtranspose}}{\NPRadjointVar}
\right\rangle.
\end{align}
This implies that we can also write an integral representation for the adjoint operator:
\begin{align} \label{eq:adjointIntegralRepresentation}
\NPRFrechetAdj{\NPRnormedRateMatrix}{\NPRadjointVar}
=
\NPRFrechetDerivative{\NPRnormedRateMatrix^{\NPRtranspose}}{\NPRadjointVar}
 = \NPRtimeValue
\int_0^1
e^{(1-s)\NPRtimeValue\NPRnormedRateMatrix^{\NPRtranspose}}
\NPRadjointVar\,
e^{s\NPRtimeValue\NPRnormedRateMatrix^{\NPRtranspose}}
\dd s.
\end{align}
Substituting the block-diagonal decomposition $\NPRnormedRateMatrix = \NPRRotation \NPRBlkDiag \NPRRotation^{-1}$ from Equation~\eqref{npr.eq:blockdiagonalrepresentation} into this last equation,
we obtain:
\begin{align}\label{eq:rotationExplicitAdjointOperator}
\NPRFrechetAdj{\NPRnormedRateMatrix}{\NPRadjointVar}
 = \NPRtimeValue   \NPRRotation^{-\NPRtranspose} \Biggr[
\int_0^1
e^{(1-s)\NPRtimeValue\NPRBlkDiag^{\NPRtranspose}}
\NPRRotation^{\NPRtranspose}\NPRadjointVar \NPRRotation^{-\NPRtranspose}\,
e^{s\NPRtimeValue\NPRBlkDiag^{\NPRtranspose}} \dd s\Biggr] \NPRRotation^{\NPRtranspose}.
\end{align}
\paragraph{Closed-form adjoint Fr\'echet derivative}
Consider now the likelihood representation
introduced in Equation~\eqref{NPR.eq:characterizationTDLPrePostOrder}, $\NPRpreOrder{\NPRchildNode}^{\NPRtranspose}
\mathbf{P}(\NPRbranchLength{\NPRchildNode})
\NPRpostOrder{\NPRchildNode}$, where $\NPRpreOrder{\NPRchildNode}$ and $\NPRpostOrder{\NPRchildNode}$ are the partial likelihood vectors at the beginning and end of an edge of length $\NPRbranchLength{\NPRchildNode}$.
Then, the adjoint of the transition matrix $\mathbf{P}(\NPRbranchLength{\NPRchildNode})$ is the outer product $\NPRpreOrder{\NPRchildNode} \NPRpostOrder{\NPRchildNode}^{\NPRtranspose}$.
Let $\NPRrotatedPreOrder_{\NPRchildNode}^{\vphantom{\NPRtranspose}} = \NPRRotation^{\NPRtranspose} \NPRpreOrder{\NPRchildNode}^{\vphantom{\NPRtranspose}} $ and $\NPRrotatedPostOrder_{\NPRchildNode}=\NPRRotation^{-1} \NPRpostOrder{\NPRchildNode}$ be the eigenbasis versions of the two partial likelihood vectors, and let $\NPRtransitionMatrixAdjoint{\NPRchildNode} = (\NPRRotation^{\NPRtranspose} \NPRpreOrder{\NPRchildNode}^{\vphantom{\NPRtranspose}})(\NPRpostOrder{\NPRchildNode}^{\NPRtranspose} \NPRRotation^{-\NPRtranspose}) = \NPRrotatedPreOrder_{\NPRchildNode}^{\vphantom{\NPRtranspose}} \NPRrotatedPostOrder_{\NPRchildNode}^{\NPRtranspose}$ be the eigenbasis transition matrix adjoint.
Then, we can apply the adjoint operator from Equation~\eqref{eq:rotationExplicitAdjointOperator} to $\NPRtransitionMatrixAdjoint{\NPRchildNode}$ and notice that the integral between brackets can be rewritten as
\begin{align}  \label{NPR.eq:fullBlockIntegral}
\int_0^1
e^{(1-s)\NPRbranchLength{\NPRchildNode} \NPRBlkDiag^{\NPRtranspose}}
\NPRtransitionMatrixAdjoint{\NPRchildNode}\,
e^{s\NPRbranchLength{\NPRchildNode} \NPRBlkDiag^{\NPRtranspose}}
\dd s = \int_0^1
e^{(1-s)\NPRbranchLength{\NPRchildNode} \NPRBlkDiag^{\NPRtranspose}}
\NPRrotatedPreOrder_{\NPRchildNode}^{\vphantom{\NPRtranspose}}\NPRrotatedPostOrder_{\NPRchildNode}^{\NPRtranspose} \,
e^{s\NPRbranchLength{\NPRchildNode} \NPRBlkDiag^{\NPRtranspose}}
\dd s.
\end{align}
Since $\NPRBlkDiag$ is block diagonal, we can solve the integral by separately solving the integrals associated with each pair of blocks in $\NPRBlkDiag$.
In particular, consider two blocks indexed by $\NPRBlockOneIndex$ and $\NPRBlockTwoIndex$.
The $(\NPRBlockOneIndex,\NPRBlockTwoIndex)$ block of the matrix solving the integral above can be derived by solving
\begin{align} \label{NPR.eq:singleBlockIntegral}
\int_0^1
e^{(1-s)\NPRbranchLength{\NPRchildNode} \NPRBlkDiag_{\NPRBlockOneIndex}^{\NPRtranspose}}
\NPRtransitionMatrixAdjointBlock{\NPRchildNode}{\NPRBlockOneIndex\NPRBlockTwoIndex}\,
e^{s\NPRbranchLength{\NPRchildNode} \NPRBlkDiag_{\NPRBlockTwoIndex}^{\NPRtranspose}}
\dd s = \int_0^1
e^{(1-s)\NPRbranchLength{\NPRchildNode} \NPRBlkDiag_{\NPRBlockOneIndex}^{\NPRtranspose}}
\NPRrotatedPreOrder_{\NPRchildNode, \NPRBlockOneIndex}^{\vphantom{\NPRtranspose}}\NPRrotatedPostOrder_{\NPRchildNode ,\NPRBlockTwoIndex}^{\NPRtranspose}\,
e^{s\NPRbranchLength{\NPRchildNode} \NPRBlkDiag_{\NPRBlockTwoIndex}^{\NPRtranspose}}
\dd s.
\end{align}
In Section~\ref{Supp-npr.sec:analyticalSolvesBlockIntegrals} of the Supplement, we provide closed-form solutions of each of these integrals for all the combinations of $1\times1$ and $2\times2$ blocks.
It is worth noting that all of these $(\NPRBlockOneIndex, \NPRBlockTwoIndex)$ integrals are independent of each other.
This means that parallelization opportunities arise both \textit{within} and \textit{between} edges.

As in the previous section, we can write the gradient of the likelihood $\NPRmyFunc$ by leveraging its representation from Equation~\eqref{NPR.eq:characterizationTDLPrePostOrder} and the reverse-mode chain rule:
\begin{align}  \label{NPR.eq:exactReverseBackpropCTMCGradientDerivation}
    \myPartial{\NPRmyFunc}{\NPRnormedRateMatrix}
    &= \sum_{\NPRchildNode}
    \NPRFrechetAdj{\NPRnormedRateMatrix}{\NPRpreOrder{\NPRchildNode}^{\vphantom{\NPRtranspose}} \NPRpostOrder{\NPRchildNode}^{\NPRtranspose}} \nonumber\\
    &= \sum_{\NPRchildNode} \NPRbranchLength{\NPRchildNode}   \NPRRotation^{-\NPRtranspose} \Biggr[
\int_0^1
e^{(1-s)\NPRbranchLength{\NPRchildNode} \NPRBlkDiag^{\NPRtranspose}}
\NPRtransitionMatrixAdjoint{\NPRchildNode}\,
e^{s\NPRbranchLength{\NPRchildNode} \NPRBlkDiag^{\NPRtranspose}} \dd s\Biggr] \NPRRotation^{\NPRtranspose} \nonumber\\
&= \NPRRotation^{-\NPRtranspose} \Biggr[\sum_{\NPRchildNode} \NPRbranchLength{\NPRchildNode}
\int_0^1
e^{(1-s)\NPRbranchLength{\NPRchildNode} \NPRBlkDiag^{\NPRtranspose}}
\NPRtransitionMatrixAdjoint{\NPRchildNode}\,
e^{s\NPRbranchLength{\NPRchildNode} \NPRBlkDiag^{\NPRtranspose}} \dd s\Biggr] \NPRRotation^{\NPRtranspose}.
\end{align}
The last line clarifies that the accumulation of the adjoints happens in the eigenbasis, and only one change-of-basis (two matrix multiplications) is needed to transform the gradient back to the original basis.

Therefore, evaluating Equation~\eqref{NPR.eq:exactReverseBackpropCTMCGradientDerivation} requires $\mathcal{O}(\NPRnstates^3 + \NPRndata\NPRnstates^2)$ operations where:
\begin{itemize}
    \item $\NPRnstates^3$ represents the one-time eigendecomposition of $\NPRnormedRateMatrix$ and the two final matrix multiplications for the change-of-basis, while
    \item $\NPRndata\NPRnstates^2$ is due to computing all the pairs $(\NPRpreOrder{\NPRchildNode}, \NPRpostOrder{\NPRchildNode})$, which is inherently a serial operation, and solving the block integrals, which are fully parallelizable as noted above.
\end{itemize}
This total cost coincides with the cost of the approximate gradient, that is, about two likelihood evaluations, plus the block integral solves.
Regarding memory, we note that during the computation of each $(\NPRpreOrder{\NPRchildNode}, \NPRpostOrder{\NPRchildNode})$, the quantities stored should actually be their eigenbasis versions $(\NPRRotation^{\NPRtranspose} \NPRpreOrder{\NPRchildNode}, \NPRRotation^{-1} \NPRpostOrder{\NPRchildNode})$.
This implies that the matrix $\NPRtransitionMatrixAdjoint{\NPRchildNode}$ need never be materialized, and we can pass just the subsets of the two vectors $(\NPRRotation^{\NPRtranspose} \NPRpreOrder{\NPRchildNode}, \NPRRotation^{-1} \NPRpostOrder{\NPRchildNode})$ to the integral solvers following Equation~\eqref{NPR.eq:singleBlockIntegral}.
The latter two columns of Table~\ref{NPR.tab:complexity} summarize the costs and caching steps presented above.

\subsubsection{Choice between exact and approximate gradients and noisy observations}\label{NPR.sec:choosingGradient}
\paragraph{Exact versus approximate} The exact and approximate gradients provide complementary approaches whose relative utility should be assessed in terms of overall HMC efficiency.
The exact gradient has a higher cost for each leapfrog step (see Table~\ref{NPR.tab:complexity}), while the approximate gradient, if inaccurate, may result in poorer Hamiltonian trajectories, leading to smaller step sizes and lower acceptance probabilities.
Thus, the approximate gradient is preferable only when the speed-up from cheaper gradient evaluations outweighs any loss in proposal quality.
Following the results in \citet{didier_surprising_2024}, one should consider the approximate gradient especially when the edge lengths are short, since in that case the first-order approximation is expected to be accurate.

\paragraph{Noisy observations and gradients} Consider now the framework for noisy observations introduced in Section~\ref{NPR.sec:likelihoodEvaluation}.
In that construction, observation noise is represented by additional emission edges whose transition matrices do not depend on time or on the CTMC rate matrix.
These edges modify only the partial likelihood vectors passed through the graph.
They do not introduce new matrix exponential derivatives, nor do they change the chain-rule structures in Equations~\eqref{NPR.eq:approximateForwardCTMCGradientDerivation} and~\eqref{NPR.eq:exactReverseBackpropCTMCGradientDerivation}.
Consequently, the asymptotic cost of both the exact and approximate gradients is unchanged under this more general scenario.

\section{Adaptation to phylogenetic and phylogeographic inference} \label{sec:NPRPhylogeneticsApplication}
Phylogenetics and phylogeography are fields that heavily rely on CTMCs, and represent the primary application motivating the methods developed in this work.
The distinguishing characteristic of these fields is that the DAG underlying the likelihood is a \emph{binary tree}.
We begin by reviewing the fundamental building blocks of Bayesian phylogenetics (Section~\ref{NPR.sec:introToBayesianPhylogenetics}) and motivating the use of the exact and approximate gradient strategies described above (Section~\ref{NPR.sec:HMCbayesianPhylo}).
We also explicitly discuss the special case of multi-site data, which is typical in phylogenetic inference, and why it strongly benefits from our contributions (Section~\ref{NPR.sec:multi-site}).
We then benchmark the proposed methods against the main competitors and validate them on synthetic and real data examples (Section~\ref{NPR.sec:results}).

\subsection{Introduction to Bayesian phylogenetics} \label{NPR.sec:introToBayesianPhylogenetics}

Phylogenetic inference relies on data with tree-structured dependencies.
These dependencies are encoded by a graphical object called a phylogeny.
A \textbf{phylogeny} is defined by the pair $\mathcal{F} = \langle \tau, \mathcal{B} \rangle$ consisting of an acyclic, connected topology $\tau = (\nodesSet, \edgesSet)$ and a collection $\mathcal{B}$ of branch lengths, where $\nodesSet$ is a set of nodes and $\edgesSet \subseteq \nodesSet \times \nodesSet$ is a set of branches.
It is also common to assume that the phylogeny is rooted and binary.
This means that, defining the \textit{degree} of a node as the number of edges directly connected to it and denoting by $\NPRnTaxa$ the number of degree-one nodes (tree tips), the phylogeny contains a single node with degree two, the root, and $\NPRnTaxa - 2$ degree-three internal nodes.
The data-generating process consists of a CTMC that originates at the tree root, evolves along the tree in accordance with its induced conditional independence structure, and is finally observed at the tree tips.

\paragraph{Inference target} Let $\NPRunnormedRateMatrix(\NPRparameterVector)$ represent the rate matrix of the CTMC evolving along an unknown tree $\NPRphylogeny$.
Given tip observations $\mathbf{\NPRdataVector}= (\NPRdata{1}, \NPRdata2, \dots, \NPRdata{\NPRnTaxa})$, we seek the posterior distribution
\begin{align} \label{npr.Eq:phyloPosteriorDistribution}
    \condprob{\NPRparameterVector,\NPRphylogeny}{\NPRdataVector} \propto \condprob{\NPRdataVector}{\NPRphylogeny,\NPRunnormedRateMatrix(\NPRparameterVector)} \prob{\NPRparameterVector}\prob{\NPRphylogeny},
\end{align}
where $\condprob{\NPRdataVector}{\NPRphylogeny,\NPRunnormedRateMatrix(\NPRparameterVector)}$ is the \textit{tree data likelihood}, while $\prob{\NPRphylogeny}$ and $\prob{\NPRparameterVector}$ are the priors on the phylogeny and the CTMC parameters.

\subsection{HMC and gradients} \label{NPR.sec:HMCbayesianPhylo}
We employ a Metropolis-within-Gibbs MCMC framework to sample from the full posterior distribution in Equation~\eqref{npr.Eq:phyloPosteriorDistribution} \citep{metropolis_equation_1953, geman_stochastic_1984}.
Within this framework, we partition the parameters into two blocks: the CTMC parameters, $\NPRparameterVector$, and a second block containing all other parameters.
While we adopt standard random-walk kernels to propose the second block of parameters, consistent with the rich and established literature in Bayesian phylogenetics \citep{hassler_data_2023}, we adopt HMC sampling for the CTMC parameters following the details we presented in Section~\ref{NPR.sec:Inference}.

\subsubsection{Random phylogeny and gradient caching}
When the phylogeny is treated as unknown, MCMC repeatedly proposes changes to the tree topology or branch lengths and re-evaluates the tree likelihood.
Our exact gradient algorithm naturally caches the relevant post-order and pre-order vectors in the eigenbasis (see Table~\ref{NPR.tab:complexity}), so single-branch updates can be evaluated cheaply as the $\mathcal{O}(\NPRnstates)$ weighted inner product $(\NPRRotation^{\NPRtranspose}\NPRpreOrder{\NPRchildNode})^{\NPRtranspose} e^{t^*\NPRBlkDiag}(\NPRRotation^{-1}\NPRpostOrder{\NPRchildNode})$, where $t^*$ is the newly proposed branch-length value.

\subsection{Multi-site data} \label{NPR.sec:multi-site}
\paragraph{Multi-site likelihood} In contrast to phylogeography, phylogenetic inference often relies on observations at multiple sites.
The most natural scenario is when the data are DNA sequences, where each site represents the observed nucleotide at a fixed position (``site'') in the sequence.
Let $\NPRnumberOfAlignmentsSites$ be the number of sites.
The data are then a collection of vectors
$(\NPRdataVector^{(1)}, \NPRdataVector^{(2)}, \dots, \NPRdataVector^{(\NPRnumberOfAlignmentsSites)})$, where $\mathbf{\NPRdataVector}^{(\NPRalignmentIndex)}= (\NPRdata{1}^{(\NPRalignmentIndex)}, \NPRdata{2}^{(\NPRalignmentIndex)}, \dots, \NPRdata{\NPRnTaxa}^{(\NPRalignmentIndex)})$, $\NPRalignmentIndex = 1, \dots, \NPRnumberOfAlignmentsSites$.
Most of the phylogenetic literature builds on the hypothesis that the site-specific CTMCs evolve independently along the tree.
Under this independence assumption, and when the sites share the same rate matrix $\NPRunnormedRateMatrix(\NPRparameterVector)$, the tree data likelihood $\condprob{\NPRdataVector^{(\NPRalignmentIndex)}}{\NPRphylogeny,\NPRunnormedRateMatrix(\NPRparameterVector)}$ needs to be evaluated at each site, so that the computational complexity scales as
\begin{align}
    \mathcal{O}(\NPRnstates^3 + \NPRnumberOfAlignmentsSites \NPRndata \NPRnstates^2) \approx  \mathcal{O}(\NPRnumberOfAlignmentsSites \NPRndata \NPRnstates^2),
\end{align}
where $\NPRnstates^3$ corresponds to the one-time rate matrix eigendecomposition, while $\NPRnumberOfAlignmentsSites \NPRndata \NPRnstates^2$ sums up the cost of the post-order traversals at each site.

\paragraph{Multi-site gradients}
For the approximate gradient, each site requires the same two traversals as in the single-site case, followed by the accumulation of branch-wise outer products.
The cost therefore scales linearly with the number of sites.
The exact gradient has the same site-wise traversal cost, but its additional block-integral solves do not scale with the number of sites: for each branch, finite-time transition matrix adjoints are first accumulated across sites, and only their sum is passed to the integral in~\eqref{NPR.eq:fullBlockIntegral}.
Thus, the extra cost of the exact method is amortized as $\NPRnumberOfAlignmentsSites$ grows, and the exact gradient approaches the cost of the approximate gradient in the many-site regime.
Both methods then cost approximately two full tree data likelihood evaluations,
\begin{align}
    \mathcal{O}(\NPRnstates^3 + \NPRnumberOfAlignmentsSites \NPRndata \NPRnstates^2) \approx  \mathcal{O}(\NPRnumberOfAlignmentsSites \NPRndata \NPRnstates^2).
\end{align}
Competing methods scale as $\mathcal{O}(\NPRnumberOfAlignmentsSites \NPRndata \NPRnstates^3)$, so the scaling gains from the two proposed approaches become especially important in the multi-site setting.

\paragraph{Across-site heterogeneity and exact gradients}
A standard way to model across-site heterogeneity is to assign each site to one of a finite number of evolutionary-rate categories \citep{yang_maximum_1993}.
The exact gradient can exploit this structure by accumulating the branch-specific finite-time transition matrix adjoints $\NPRtransitionMatrixAdjoint{}$ within each category before applying the block-integral operator (Section~\ref{NPR.sec:exactAdjointGradient}).
Consequently, the additional integral-solve cost scales with the number of rate categories rather than with the number of sites.

\subsection{Results} \label{NPR.sec:results}
In this section, we first benchmark the proposed gradient computation methods against competing methods.
We then assess statistical recovery in a simulation study, and finally illustrate the proposed model on two real-world datasets.
All the following analyses were run using the popular open-source software \textsc{Beast X} \citep{baele_beast_2025}, with support from the high-performance computing environment \textsc{Beagle} \citep{ayres_beagle_2019}.
The simulated data were generated using $\pi$\textsc{Buss} \citep{bielejec_buss_2014}.
The code and data to reproduce our analysis are available in the public GitHub repository \url{https://github.com/suchard-group/NonParametricModelingofCTMCs}.

\subsubsection{Benchmarking} \label{NPR.sec:benchmarking}
We first isolate the computational question: how does the wall-clock time of one likelihood-gradient evaluation scale with the number of CTMC states and with the number of tree tips?
We compare the proposed exact and approximate spectral gradients, implemented in \textsc{Beast X}, with three alternatives: \texttt{torchtree} \citep{fourment_torchtree_2026}, a PyTorch-based autodifferentiation framework; \texttt{treeflow} \citep{swanepoel_treeflow_2022}, a TensorFlow-based autodifferentiation framework; and \texttt{PhyloGrad} \citep{lieser_phylograd_2025}, a dedicated and optimized reverse-mode implementation for reversible models.

\paragraph{Conditions} The benchmark uses fixed binary trees with $\NPRndata \in \{100,1000,10000\}$ tips and varies the number of CTMC states over $\NPRnstates \in \{8,16,32,64,128,256\}$.
Within each configuration, all methods use the same tree, branch lengths, and observed tip states.
Timings are obtained
using a single CPU thread on UCLA's Hoffman2 high-performance computing cluster.
We consider both reversible and non-reversible rate matrices.
\texttt{PhyloGrad} is included only in the former case because it assumes reversibility.
For \texttt{treeflow}, we use a numerically rescaled pruning implementation with TensorFlow batch matrix exponentials, since \texttt{treeflow} pruning underflows for large trees.

\begin{figure}[!htbp]
    \centering
    \includegraphics[width=\linewidth]{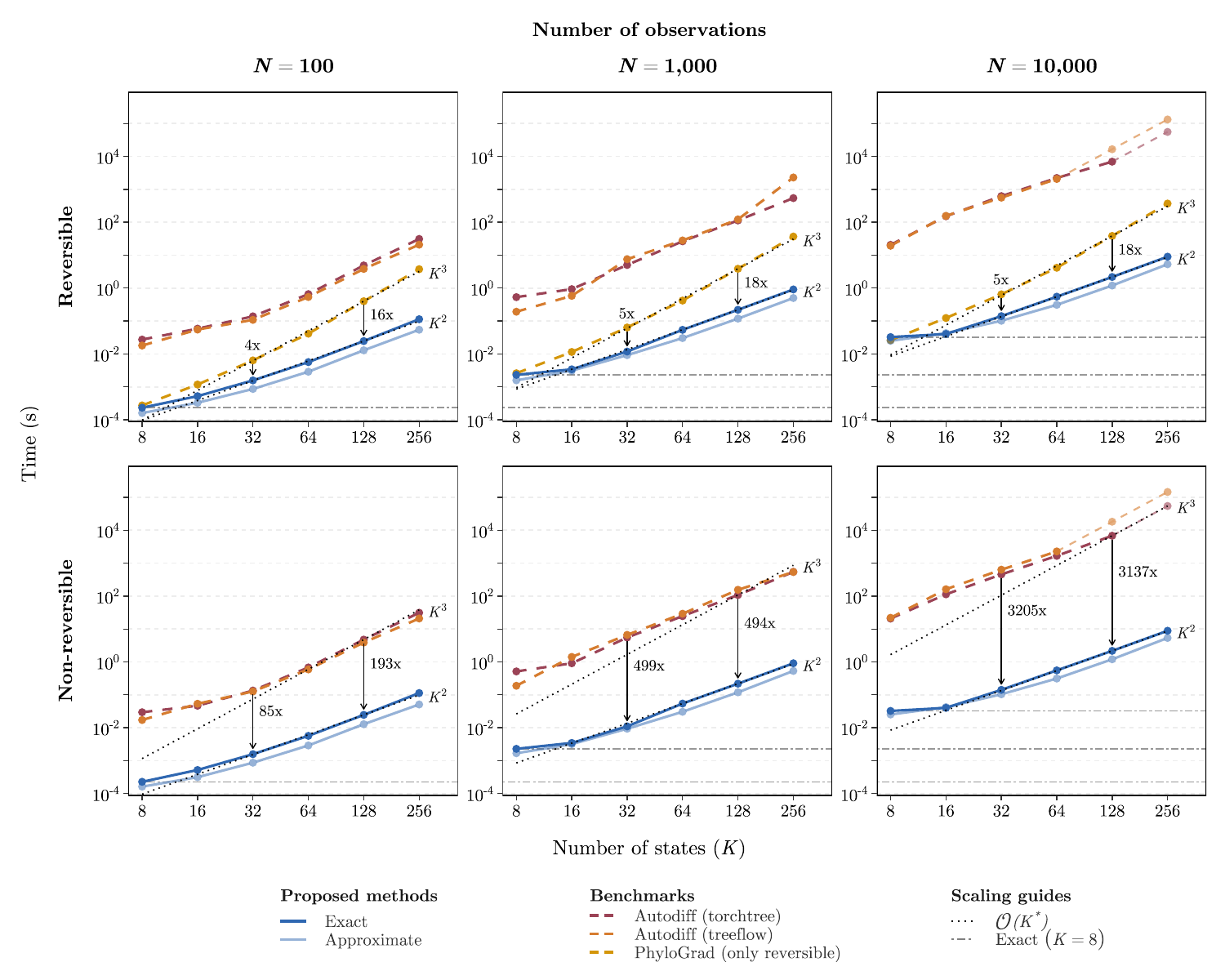}
    \justify
    \captionsetup{font={stretch=1}}
    \caption{\textbf{Gradient timing benchmark.}
    Median wall-clock time per gradient evaluation (seconds, log scale), after warm-up, as a function of the number of CTMC states $\NPRnstates$ for the exact and approximate gradients and three competitor frameworks.
    Rows correspond to reversible (top) and non-reversible (bottom) rate matrices; columns correspond to sample sizes $\NPRndata \in \{100, 1000, 10000\}$.
    Dotted grey lines show $\mathcal{O}(\NPRnstates^2)$ and $\mathcal{O}(\NPRnstates^3)$ reference slopes anchored at the $\NPRnstates = 128$ values respectively for the exact gradient and its fastest competitor.
    Vertical arrows show speed-ups between the exact gradient and its fastest competitor at $\NPRnstates = 32$ and $\NPRnstates = 128$.
    Horizontal dash-dot lines mark the timings of the exact gradient at $\NPRnstates = 8$ in each panel, providing a visual reference for the near-linear increase in runtime as $\NPRndata$ increases across columns.
    Transparent dots represent extrapolated cases where the run did not finish, for example, because of memory exhaustion; extrapolations use the corresponding theoretical scaling in $\NPRnstates$ anchored at the largest completed run.}
    \label{NPR.fig:gradientBenchmark}
\end{figure}

Figure~\ref{NPR.fig:gradientBenchmark} reports median wall-clock times after warm-up.
For configurations with median runtimes below $100$ seconds, timings are based on $200$ warm-up gradient evaluations followed by $200$ timed evaluations; for slower configurations, the number of evaluations was reduced to keep the benchmark computationally feasible.
Transparent points in the figure denote extrapolated configurations that did not complete, for example because of memory exhaustion, and are shown only for visual continuity.

\paragraph{Results} Across the measured configurations, the proposed exact and approximate gradients are consistently faster than the competing methods.
For example, at $\NPRnstates=128$, the exact gradient is tens of times faster than the fastest measured competitor in the reversible case.
In the non-reversible case, where \texttt{PhyloGrad} is not applicable, the corresponding speed-ups shift to hundreds and thousands as the number of tips increases from $100$ to $10000$.
The approximate gradient has the same qualitative scaling as the exact gradient, but is typically about $1.5$--$2\times$ faster at larger state dimensions because it avoids the exact method's additional block-integral solves.

The empirical scaling curves are consistent with the theoretical complexity analysis in Table~\ref{NPR.tab:complexity}.
The exact and approximate gradients follow an approximately quadratic scaling pattern in $\NPRnstates$ over the benchmark range.
This trend also appears consistently in scenarios with $\NPRndata < \NPRnstates$, meaning the cubic cost of the eigendecompositions is well amortized even in these cases.
By contrast, the competing analytical reverse-mode and autodifferentiation strategies approach the cubic reference slope as $\NPRnstates$ increases.
It is worth noting that, for the largest tree sizes, the autodifferentiation curves (\texttt{torchtree} and \texttt{treeflow}) appear to flatten.
This should not be interpreted as evidence of improved asymptotic scaling: these implementations involve substantial framework and tensor-management overheads, whose large constants can dominate wall-clock time and partially obscure the expected state-space scaling over the measured range.

\subsubsection{Simulation} \label{NPR.sec:simulation}
To assess the performance of the proposed methods for parameter estimation, we conduct a simulation study.
Based on \citet{faria_toward_2011}, we begin by considering a CTMC describing the transmission process of the rabies virus infecting $17$ host bat species.
Each pair of species is associated with their genetic distances.

\begin{figure}[ht]
    \centering
    \includegraphics[width=0.55\linewidth]{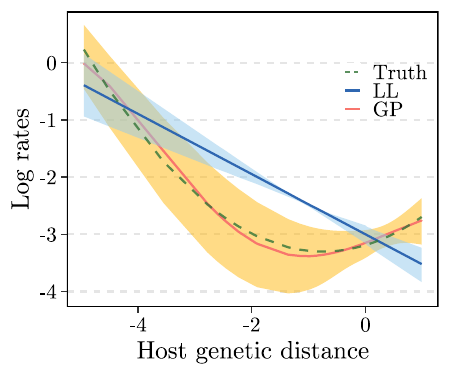}
    \caption{\textbf{Simulation studies.}
    The figure compares log-linear (LL) and Gaussian process (GP) models in recovering CTMC log-rates from simulated data where true log-rates (dashed line) follow a quadratic function of host genetic distance.
    Solid lines show posterior median normalized log-rates; shaded regions represent 95\% highest posterior density intervals (HPDIs).
    Only slope uncertainty is depicted due to normalization.
    }
    \label{NPR.fig:simulation}
    \end{figure}

For simulation purposes, we set the log-rates of the CTMC equal to a quadratic function of those genetic distances as shown in Figure~\ref{NPR.fig:simulation} (dashed line).
Next, we select a phylogenetic tree sampled from the posterior distribution given aligned genomic sequences obtained in \citet{faria_toward_2011}, containing $372$ tips and spanning an evolutionary history of approximately 270 years.
Subsequently, we simulate a single alignment (one observation per tip) by evolving the CTMC along the tree branches, starting from the root.
The evolutionary rate, representing the average ``speed'' of evolution, is fixed at $2\%$ per year.
This corresponds to an expected total of roughly five transitions along the path from the root to the most recent tip.

To analyze the simulated data, we employ a GP prior on the log-rates, using a squared exponential kernel with the host-species genetic distances as a covariate.
We place independent exponential priors with rate $1$ on the kernel hyperparameters (the marginal scale and the length-scale).
As shown in Figure~\ref{NPR.fig:simulation} (red line), the method recovers the underlying nonlinear log-rate function.
For comparison, we also ran the analysis using a log-linear model (LL, blue line).
Quantitatively, across posterior draws, the GP achieves a median root mean square error (RMSE) of $0.170$ ($95\%$ highest posterior density interval (HPDI) $0.039$--$0.311$), versus $0.487$ ($95\%$ HPDI $0.404$--$0.589$) for the LL model---a $\approx65\%$ reduction (LL $\approx2.9\times$ larger), indicating superior accuracy.
Moreover, pointwise $95\%$ HPDI coverage of the true log-rate function is $100\%$ for the GP and $28.68\%$ for the log-linear model, with average HPDI widths of $0.63$ and $0.39$, respectively; thus, the GP bands are slightly wider, as expected, but substantially better calibrated.

We note that updating the hyperparameters entails a high computational cost.
Indeed, when the kernel is modified, the GP covariance matrix also changes, requiring operations with cubic-time complexity in its dimension (e.g., inversion or Cholesky decomposition).
Since the Gaussian density must be evaluated for $\NPRnstates^2 - \NPRnstates$ elements, the overall computational complexity increases to ${\cal O}(\NPRnstates^6)$.
Fortunately, the hyperparameters are low-dimensional, and it therefore suffices to update them only occasionally.
A profiling analysis of our simulation indicates that this cost is negligible when $\NPRnstates = 17$.
However, as the state space grows, this cost may become non-trivial. Still, it should be compared with the cost of the likelihood and gradient evaluations, which both scale with the number of states and the number of observations—the latter of which must also increase to provide sufficient information for all analyzed states.

\begin{figure}[ht]
    \begin{subfigure}{0.45\textwidth}
        \centering
        \includegraphics{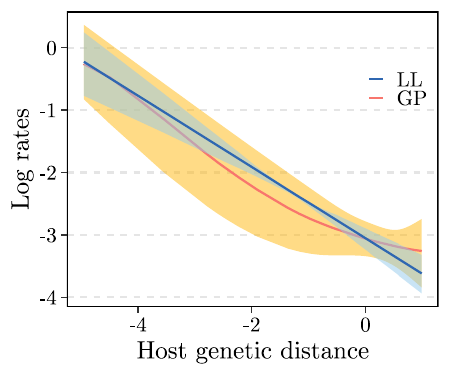}
        \caption{Rabies viruses in bats}
        \label{NPR.fig:rabiesGpHostDistanceE}
    \end{subfigure}
    \hspace{0.02\textwidth}
    \begin{subfigure}{0.45\textwidth}
        \centering
        \includegraphics{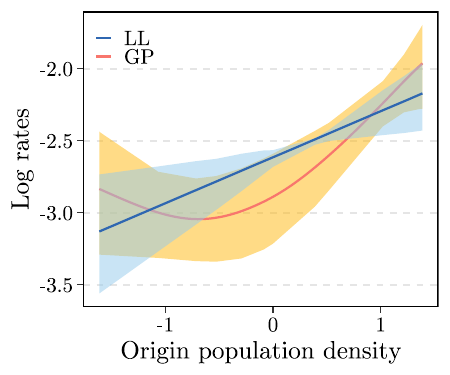}
        \caption{Global flu}
    \label{NPR.fig:fluLogRates}
    \end{subfigure}
    \caption{\textbf{Data examples: log-rates vs predictors.} Panel (a) shows the effect of cross-species genetic distances on rabies virus transmission rates; panel (b) shows the effect of origin country population density on global flu transmission.
    Solid lines represent posterior median log-rates inferred under log-linear (LL) and Gaussian process-based (GP) models; shaded areas show 95\% HPDIs.
    Only slope uncertainty is depicted due to normalization.
    }
\end{figure}

\subsubsection{Bat rabies viruses in North America}  \label{NPR.sec:realDataBat}

After validating our method on simulated data, we apply it to two real-world examples.
We start by reproducing the analysis performed in \citet{faria_simultaneously_2013} of a dataset composed of $372$ rabies virus samples collected from $17$ species of infected bats in North America between 1997 and 2006 \citep{streicker_host_2010}.
This study reconstructs the evolutionary history of the rabies viruses using nucleoprotein gene sequences, sample locations, and the host species of each infected bat.

    Moreover, the authors investigate the transmission dynamics between different bat species to identify potential facilitating factors.
To do so, they model the \emph{host species} associated with each viral lineage as a CTMC evolving along the branches of the viral phylogeny.
The state space of the CTMC therefore corresponds to the $17$ bat species observed in the dataset; transitions of the CTMC represent cross-species transmission events; and they adopt cross-species genetic distance as a covariate.
Using a log-linear model, they conclude that cross-species genetic distance influences transmission: the more genetically different two bat species are, the less likely a transmission event is between them.

\begin{figure}[!htbp]
    \centering
    \includegraphics[width=\linewidth]{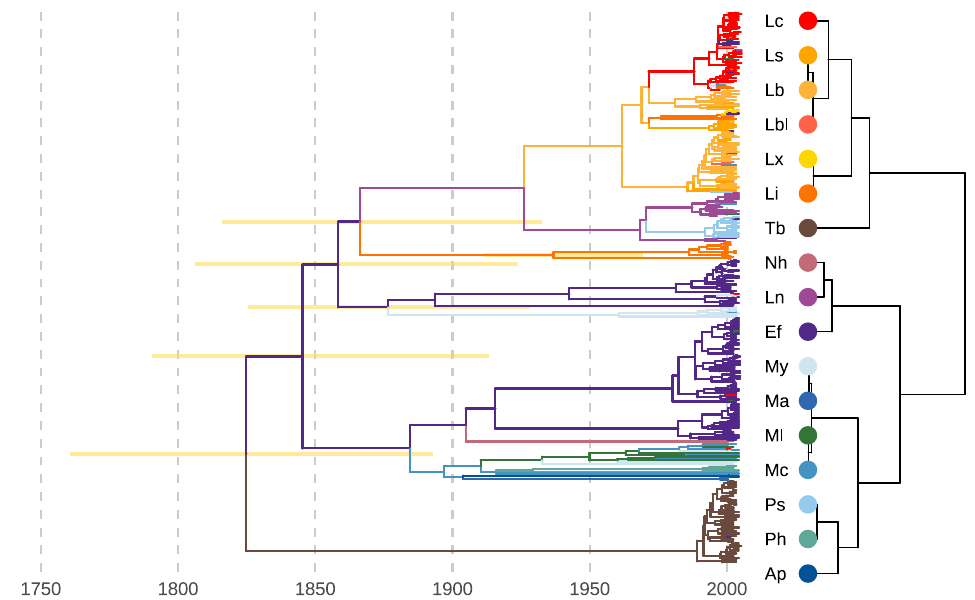}
    \justify
    \caption{\textbf{Bat Rabies data example.} The viral tree (left) shows the rabies virus evolution in North American bats.
    Yellow bars represent 95\% HPDIs for node ages; branch colors identify the most probable host bat species.
    The bat species tree (right) orders bat species by genetic distance using Ward's hierarchical clustering algorithm \citep{ward_jr_hierarchical_1963}.
    Color symmetry between trees suggests rabies transmission occurs preferentially among genetically similar bat species.
    }
    \label{NPR.fig:batRabiesTree}
\end{figure}

We revisit these data, relaxing the log-linear assumption.
Specifically, we assign to the infinitesimal log-rates a GP prior with a squared exponential kernel using the host genetic distance as a covariate.
Moreover, we specify two exponential priors with rates $1$ and $2$, respectively, for the kernel scale and length hyperparameters.
To avoid overfitting, we also constrain the length parameter to be at least one.
As Figure~\ref{NPR.fig:rabiesGpHostDistanceE} shows, our results support the hypothesis that genetic distance has a log-linear effect on the infinitesimal rates.
The relationship between this covariate and rabies virus evolutionary history can also be visualized by comparing the two trees shown in Figure~\ref{NPR.fig:batRabiesTree}.
The left panel shows a viral tree produced according to the maximum clade credibility (MCC) criterion \citep{Baele_HIPSTR_2025}.

This method selects the single tree that best represents the most well-supported clades from the full distribution of sampled trees, where a clade can be defined as a subset of the complete tree including a common ancestor and all of its descendants.

Branches are colored according to the bat species most likely to harbor the virus lineage during the corresponding period.

On the right, a dendrogram illustrates the hierarchical clustering of bat species based on cross-species genetic distances, with groupings determined by Ward's method \citep{ward_jr_hierarchical_1963}, that is, by minimizing the within-cluster variance.
It is apparent that the two trees have a symmetric color-induced structure, suggesting a tendency for infection events to occur more frequently across more closely related species.

\subsubsection{Global influenza and human transportation}  \label{NPR.sec:realDataAirco}
As a second example, we use data on the global spread of seasonal H3N2 influenza from 2002 to 2007.
The data consist of hemagglutinin sequences and sample locations from 1,441 human-sampled viral isolates,
and were originally analyzed by \citet{lemey_unifying_2014} to study the annual spread of the virus across 14 global air communities.
The authors' original work used a log-linear model to guide the spatial dispersion process, leveraging covariates such as airplane seat numbers and origin country population densities.

Focusing on origin country population densities, we aim to relax the log-linear assumption by modeling the covariate effect using a GP.
We adopt a squared-exponential kernel and assign two exponential priors with rates $1$ and $2$, respectively, to the marginal scale and length-scale hyperparameters.
To avoid overfitting, we constrain the length parameter to be greater than $1$.

Figure~\ref{NPR.fig:fluLogRates} compares the inferred log-rates from both the log-linear and the GP models.
The results suggest that the log-linear assumption is too restrictive, as it appears to overestimate the effect of low population densities while underestimating the effect of higher ones.
\begin{figure}[!htbp]
    \includegraphics[width=\linewidth]{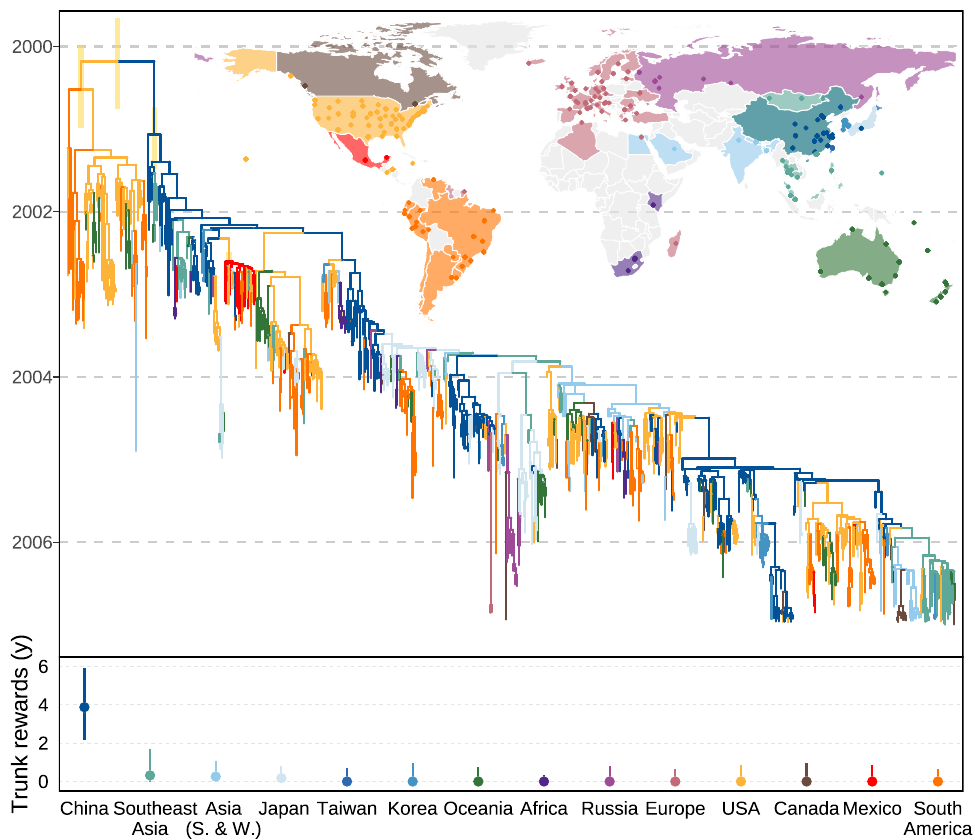}
    \justify
    \caption{\textbf{Global influenza data example.}
    Maximum clade credibility (MCC) tree showing H3N2 influenza evolutionary history.
    Yellow bars on internal nodes represent 95\% HPDIs for node ages.
    Branch colors indicate distinct air communities (map, top-right; dots represent analyzed airports), with each color showing the air community where the GP model predicts most time is spent.
    The bottom plot shows trunk rewards (cumulative time in years the CTMC spends in each state along the trunk, defined as external branches facing top-right) based on the GP model, with error bars representing 95\% HPDIs.
    }
    \label{NPR.fig:fluTree}
\end{figure}
Figure~\ref{NPR.fig:fluTree} also shows the MCC tree reconstructing the evolutionary history of H3N2 influenza, where each branch is colored by the air community in which the corresponding lineage spent the most time.
A quick inspection of the tree suggests that China and, in general, countries in southern Asia were potential recurrent sources of infection.
Indeed, their corresponding colors dominate the tree trunk, which we define here as the sequence of external branches facing the top-right of the tree plot.
To test this hypothesis, we compute the reward associated with each air community along the tree trunk; that is, the total expected time the CTMC spends in each of the states along the tree trunk \citep{minin_counting_2007}.
The dot plot at the bottom of Figure~\ref{NPR.fig:fluTree} shows the posterior median trunk rewards associated with each air community using the GP model.
In particular, the figure shows that the highest median trunk reward is associated with China (around 4 years between 2000 and 2007), confirming the aforementioned hypothesis.

\section{Discussion}\label{sec:NPRDiscussion}
This work introduces a flexible framework for analyzing the influence of external covariates on the jump process of CTMCs, particularly in high-dimensional state spaces where direct parametrization is challenging.
Our primary contribution lies in relaxing the restrictive log-linear assumption prevalent in existing models by proposing a GP prior on a positive transformation of the infinitesimal rate matrix entries, modeled as a random function of covariates.
This formulation allows transition rates to vary flexibly with covariates, thereby capturing nonlinear relationships while preserving an interpretable dependence between covariates and transformed rates.
Such flexibility is particularly useful in contexts where the structure of the underlying jump process is poorly understood.
A second contribution is computational.
Flexible priors over CTMC rates are useful only if posterior inference remains feasible.
For partially observed CTMCs, this is difficult because the likelihood depends on matrix exponentials, and HMC requires repeated differentiation of these matrix exponentials.
We address this bottleneck with two complementary gradient strategies.
The first uses an approximate matrix-exponential derivative within an HMC scheme with surrogate trajectories, so that the proposal mechanism is computationally cheap while the Metropolis--Hastings correction preserves the exact posterior as the stationary distribution.
The second derives an exact reverse-mode gradient based on closed-form adjoint Fr\'echet derivatives.
Both approaches exploit the block-diagonal eigendecomposition of the rate matrix and reduce the dominant gradient cost from $\mathcal{O}(\NPRndata\NPRnstates^3)$ to $\mathcal{O}(\NPRnstates^3+\NPRndata\NPRnstates^2)$.

We implement both approaches in \textsc{Beast X}, a widely used software for Bayesian phylogenetic and phylogeographic inference.
In a benchmark study, we timed our gradient evaluations for likelihoods on binary trees with up to $10{,}000$ tree tips and $256$ CTMC states, using reversible and non-reversible rate matrices.
The results show that these computations achieve empirical quadratic scaling over the tested range of $\NPRnstates$, even in scenarios where $\NPRnstates > \NPRndata$, e.g., when $\NPRnstates=256$ and $\NPRndata=100$.
Alternative existing methods relying on autodifferentiation scale approximately cubically and, even more importantly, suffer from substantial constant factors, making them hundreds to thousands of times slower.
Beyond the benchmarks, the implementation in \textsc{Beast X}
is intended to facilitate broader adoption in the Bayesian phylogenetic community and to promote the exploration of flexible covariate effects in CTMC models.
Our simulation study validates the implementation within \textsc{Beast X}, showing that the GP model can recover nonlinear log-rate relationships more accurately than its log-linear counterpart when the true relationship is nonlinear.
Two real-data examples further illustrate the framework's utility.
In the analysis of bat rabies virus transmission, relaxing the log-linear assumption with a GP prior nevertheless supported an approximately log-linear relationship between genetic distance and cross-species transmission rates.
In contrast, for global seasonal influenza (H3N2) spread, the GP prior revealed that the log-linear model systematically misestimated the effects of origin population density, overemphasizing the impact of low population densities while underestimating that of higher ones.
This example highlights how the GP framework can uncover nonlinear covariate effects that simpler models may obscure.

Future work could extend the framework to account for uncertainty in the covariates themselves, which are currently treated as fixed and known.
Incorporating covariate uncertainty may yield more robust inference, particularly in settings where the external information is noisy or incomplete.
In addition, the multi-site gradient results derived in Section~\ref{NPR.sec:multi-site} suggest several promising directions.
As the number of sites $\NPRnumberOfAlignmentsSites$ grows, the cost of the exact gradient approaches that of the approximate one, because adjoint accumulation across sites amortizes the integral-solve cost; both methods will achieve a time complexity of around $\mathcal{O}(\NPRnumberOfAlignmentsSites \NPRndata \NPRnstates^2)$.
This favorable scaling opens the door for gradient-based inference for higher-dimensional substitution models for molecular sequence data, for example, to estimate amino-acid substitution rates using large genomic alignments.
In the across-site rate-heterogeneity setting, the linear scaling of integral solves in the number of rate categories further suggests that the exact gradient could be applied to mixture-based substitution models without prohibitive overhead.

The GP prior also introduces its own computational costs.
In particular, GP hyperparameter sampling requires operations cubic in the number of off-diagonal rates, corresponding to $\mathcal{O}((\NPRnstates^2-\NPRnstates)^3) \approx\mathcal{O}(\NPRnstates^6)$ scaling in the state-space size.
Although this cost is amortized in the examples considered here, extensions to substantially larger state spaces should also consider more scalable nonparametric or semiparametric alternatives, such as sparse or structured GP approximations, spline-based models, or compact neural-network parameterizations.

In summary, the proposed framework offers a robust and scalable solution for Bayesian inference of CTMCs.
By providing a principled way to model nonlinear covariate effects and employing computationally efficient gradient evaluations, this work extends the capabilities of current models, enabling deeper insights into the dynamics of stochastic processes.

\section*{Acknowledgments} The authors thank Moritz Kraemer and Philippe Lemey for insightful discussions and suggestions.
We gratefully acknowledge support from Advanced Micro Devices, Inc.~with the donation of parallel computing resources used for this research.

\section*{Funding} This work was supported through National Institutes of Health grants U19 AI135995, R01 AI153044, and R01 AI162611.

\bibliography{references}
\end{document}

\justifying
\doublespacing
\renewcommand{\thesection}{S\arabic{section}}
\renewcommand{\thesubsection}{S\arabic{section}.\arabic{subsection}}
\renewcommand{\theequation}{S\arabic{equation}}
\renewcommand{\thetable}{S\arabic{table}}

\title{Supplement to ``Nonparametric Modeling of Continuous-Time Markov Chains with Scalable Exact and Approximate Gradients''}
\author{\normalsize \textsc{Filippo Monti, Xiang Ji, Yucai Shao, and Marc A.~Suchard}}
\date{}
\maketitle
\section{\texorpdfstring{Gradient with respect to the auxiliary parameters $\NPRparameter{\NPRelementIndices}$ in partially observed CTMCs}{Gradient with respect to the auxiliary parameters in partially observed CTMCs}}  \label{npr.sec:gradWrtAuxiliaryParameters}

Let $\NPRunnormedRateMatrix = (\NPRunnormedElement{\NPRelementIndices})_{\NPRelementIndices}$ be the infinitesimal matrix of a CTMC from which we sample $\NPRnTaxa$ observations $\NPRdataVector = (\NPRdata{1}, \NPRdata{2}, \dots, \NPRdata{\NPRnTaxa})$.
We introduce real-valued auxiliary variables $\NPRparameter{\NPRelementIndices} = g^{-1}(\NPRunnormedElement{\NPRelementIndices})$ where $g$ is a non-negative differentiable one-to-one function.
When a CTMC is partially observed, to ensure identifiability, we adopt the normalization constant
$  \NPRnormConstant
  = \sum_{\NPRelementIndicesFirst \neq \NPRelementIndicesSecond} \NPRstationaryDistribution_{\NPRelementIndicesFirst} \NPRunnormedElement{\NPRelementIndices}$, where $\NPRstationaryDistributionVector$ is a distribution vector over the CTMC states.
Then, we define a \emph{normalized} infinitesimal generator $\NPRnormedRateMatrix = (\NPRnormedElement{\NPRelementIndices})_{\NPRelementIndices}$ element-wise by $\NPRnormedElement{\NPRelementIndices} = \frac{\NPRunnormedElement{\NPRelementIndices}}{\NPRnormConstant}$.
$\NPRnormedRateMatrix$ characterizes a new CTMC in which time and infinitesimal rates are identifiable.
In this new normalized formulation, the likelihood $\NPRmyFunc$ depends directly only on the normalized rates $\NPRnormedElement{\NPRelementIndices}$.
Therefore, differentiating $\NPRmyFunc$ with respect to the auxiliary parameters $\NPRparameter{\NPRelementIndices}$ needs to account for their effect on each of the normalized rates.
We now formally derive the likelihood gradient and evaluate it when the link function $g$ is exponential.

First, we need to apply the chain rule on $\NPRmyFunc$, separating the effect of $\NPRparameter{\NPRelementIndices}$ on each $\NPRnormedElement{\NPRelementIndices}$, that is:
\begin{align}
  \myPartial{\NPRmyFunc}{\NPRparameter{\NPRelementIndices}}
    &= \sum_{\NPRchainIndices} \myPartial{\NPRmyFunc}{\NPRnormedElement{\NPRchainIndices}}
       \myPartial{\NPRnormedElement{\NPRchainIndices}}{\NPRparameter{\NPRelementIndices}}
     = \sum_{\NPRchainIndices} \myPartial{\NPRmyFunc}{\NPRnormedElement{\NPRchainIndices}} \left(
       \frac{1}{\NPRnormConstant} \myPartial{\NPRunnormedElement{\NPRchainIndices}}{\NPRparameter{\NPRelementIndices}}
       -
       \NPRunnormedElement{\NPRchainIndices} \frac{1}{\NPRnormConstant^2} \myPartial{\NPRnormConstant}{\NPRparameter{\NPRelementIndices}}
     \right) \nonumber\\
    & = \sum_{\NPRchainIndices}
          \myPartial{\NPRmyFunc}{\NPRnormedElement{\NPRchainIndices}} \frac{1}{\NPRnormConstant} \myPartial{\NPRunnormedElement{\NPRchainIndices}}{\NPRparameter{\NPRelementIndices}}
        -
        \sum_{\NPRchainIndices}
           \myPartial{\NPRmyFunc}{\NPRnormedElement{\NPRchainIndices}} \NPRunnormedElement{\NPRchainIndices} \frac{1}{\NPRnormConstant^2 } \myPartial{\NPRnormConstant}{\NPRparameter{\NPRelementIndices}}.
           \label{NPR.eq:gradientAuxiliaryInitial}
\end{align}
Noticing that $\myPartial{\NPRunnormedElement{\NPRchainIndices}}{\NPRparameter{\NPRelementIndices}} = 0$ for all $\NPRchainIndices \neq \NPRelementIndices \text{ and } \NPRdiagonalIndices$, we can re-write the first summation in Equation~\ref{NPR.eq:gradientAuxiliaryInitial} as:
\begin{align} \label{NPR.eq:gradientAuxiliaryInitialFirstSummation}
    \sum_{\NPRchainIndices} \myPartial{\NPRmyFunc}{\NPRnormedElement{\NPRchainIndices}} \frac{1}{\NPRnormConstant} \myPartial{\NPRunnormedElement{\NPRchainIndices}}{\NPRparameter{\NPRelementIndices}}
    &= \left(
      \myPartial{\NPRmyFunc}{\NPRnormedElement{\NPRelementIndices}} - \myPartial{\NPRmyFunc}{\NPRnormedElement{\NPRdiagonalIndices}}
    \right) \frac{1}{\NPRnormConstant} \myPartial{\NPRunnormedElement{\NPRelementIndices}}{\NPRparameter{\NPRelementIndices}}
    =
    \NPRsubstitutionDifferential{\NPRelementIndices} \frac{1}{\NPRnormConstant} \myPartial{\NPRunnormedElement{\NPRelementIndices}}{\NPRparameter{\NPRelementIndices}}
  \end{align}
where $\NPRsubstitutionDifferential{\NPRelementIndices}:=\left(
\myPartial{\NPRmyFunc}{\NPRnormedElement{\NPRelementIndices}} - \myPartial{\NPRmyFunc}{\NPRnormedElement{\NPRdiagonalIndices}}
\right) $ is the difference between the partial derivative with respect to a non-diagonal element and its corresponding diagonal element.
We can then insert Equation~\ref{NPR.eq:gradientAuxiliaryInitialFirstSummation} into Equation \ref{NPR.eq:gradientAuxiliaryInitial} and input the exponential link function $\NPRunnormedElement{\NPRelementIndices} = e^{\NPRparameter{\NPRelementIndices}}$ such that:
\begin{align}
    \myPartial{\NPRmyFunc}{\NPRparameter{\NPRelementIndices}}
      &=
      \left[ \NPRsubstitutionDifferential{\NPRelementIndices}
       - \left( \sum_{\NPRchainIndices}
      \myPartial{\NPRmyFunc}{\NPRnormedElement{\NPRchainIndices}} \NPRnormedElement{\NPRchainIndices}
      \right) \NPRstationaryDistribution_{\NPRelementIndicesFirst} \right] \NPRnormedElement{\NPRelementIndices},
\end{align}
recalling that $\myPartial{\NPRnormConstant}{\NPRparameter{\NPRelementIndices}} =  \NPRstationaryDistribution_{\NPRelementIndicesFirst} \NPRunnormedElement{\NPRelementIndices}
= \NPRstationaryDistribution_{\NPRelementIndicesFirst} \NPRnormConstant \NPRnormedElement{\NPRelementIndices}$.
Notice that these expressions only depend on the distribution vector $\NPRstationaryDistributionVector$, the normalized rates $\NPRnormedElement{\NPRelementIndices}$, and the gradient of $\NPRmyFunc$ with respect to each normalized rate.
We can therefore evaluate the gradient directly using the matrix exponential approximation proposed in the main text.

\section{Real block representation}\label{NPR.sec:nonsymmetricEigendecomposition}
The normalized CTMC generator $\NPRnormedRateMatrix$ is real, but it is not generally symmetric.
Consequently, its spectrum may contain complex-conjugate eigenvalue pairs.
Standard nonsymmetric eigensolvers exploit this structure by working in real arithmetic and representing complex-conjugate eigenpairs through real two-dimensional invariant subspaces \citep{golub_matrix_2013, anderson_lapack_1999}.
Recall that a subspace is \emph{invariant} under $\NPRnormedRateMatrix$ if applying $\NPRnormedRateMatrix$ to any vector in the subspace produces another vector in the same subspace.
\paragraph{From Schur form to the block representation.}
A real Schur decomposition first writes
\begin{align}
\NPRnormedRateMatrix
=
\mathbf{Q}\mathbf{T}\mathbf{Q}^{T},
\end{align}
where $\mathbf{Q}$ is orthogonal and $\mathbf{T}$ is real upper quasi-triangular \citep{golub_matrix_2013}.
The diagonal of $\mathbf{T}$ contains $1\times1$ blocks associated with real eigenvalues and $2\times2$ blocks associated with complex-conjugate pairs, while entries above these diagonal blocks may be nonzero.
Thus, the Schur form exists for every real square matrix, but it is not generally block diagonal.
The block representation used in the main manuscript is obtained by replacing this quasi-triangular representation with a basis of real invariant subspaces.
Each real eigenvalue contributes a one-dimensional invariant subspace and hence a $1\times1$ block.
Each complex-conjugate pair $\alpha \pm \mathrm{i}\beta$, with $\beta \neq 0$, contributes a two-dimensional real invariant subspace and hence a real block of the form \citep{golub_matrix_2013}
\begin{align}
\begin{pmatrix}
\alpha & \beta \\
-\beta & \alpha
\end{pmatrix}.
\end{align}
Collecting a basis for these real invariant subspaces into $\NPRRotation$ gives
\begin{align}\label{NPR.eq:blockEigenRelation}
\NPRnormedRateMatrix \NPRRotation
=
\NPRRotation \NPRBlkDiag,
\end{align}
where $\NPRBlkDiag$ is block diagonal with $1\times1$ real eigenvalue blocks and $2\times2$ real blocks for complex-conjugate pairs.
If $\NPRRotation$ is nonsingular, this implies
\begin{align}\label{NPR.eq:blockDiagonalization}
\NPRnormedRateMatrix
=
\NPRRotation \NPRBlkDiag \NPRRotation^{-1}.
\end{align}
When $\NPRRotation$ is singular or quasi-singular, standard eigensolvers attempt this recovery through back-substitution and back-transformation, with convergence diagnostics used to detect cases where the computed basis is unreliable \citep{wilkinson_handbook_2012, anderson_lapack_1999}.

\section{Analytical Solves for Block Integrals}  \label{npr.sec:analyticalSolvesBlockIntegrals}
Let
$\NPRBlkDiag = \diag \, (\NPRBlk{\NPRBlockOneIndex})_\NPRBlockOneIndex$
be a block diagonal matrix, where each block $\NPRBlk{\NPRBlockOneIndex}$ is either $1\times1$, corresponding to a real eigenvalue, or $2\times2$, corresponding to a complex conjugate eigenvalue pair in real block form.
Let $\NPRtimeValue>0$ be a branch length and $\NPRsuppFrechetDirection$ be a generic direction matrix.
In this section, we provide closed-form solutions for the integrals of the form
\begin{align} \label{NPRsupp.eq:blockDiagonal}
\int_0^1
e^{(1-s)\NPRtimeValue\NPRBlk{\NPRBlockOneIndex}}
\NPRsuppFrechetDirectionBlock{\NPRBlockOneIndex \NPRBlockTwoIndex}\,
e^{s\NPRtimeValue\NPRBlk{\NPRBlockTwoIndex}}
\dd s,
\end{align}
where $(\NPRBlockOneIndex,\NPRBlockTwoIndex)$ are the indices of two blocks in $\NPRBlkDiag$ and $\NPRsuppFrechetDirectionBlock{\NPRBlockOneIndex \NPRBlockTwoIndex}$ is the corresponding sub-block of $\NPRsuppFrechetDirection$.

\paragraph{Scalar integral notation.}
We start by introducing the notation for the scalar divided difference of the exponential as it will play a central role in the derivations~\citep{najfeld_derivatives_1995,daletskii1965integration,mathias1996chain}:
\begin{align}\label{NPR.eq:varphiDef}
\varphi(\NPRsuppDummyIndexOne,\NPRsuppDummyIndexTwo)
=
\int_0^1
e^{(1-s)\NPRsuppDummyIndexOne}
e^{s\NPRsuppDummyIndexTwo}
\dd s
=
\begin{cases}
\dfrac{e^{\NPRsuppDummyIndexOne}-e^{\NPRsuppDummyIndexTwo}}{\NPRsuppDummyIndexOne-\NPRsuppDummyIndexTwo},
&
\NPRsuppDummyIndexOne\neq\NPRsuppDummyIndexTwo,
\\[10pt]
e^{\NPRsuppDummyIndexOne},
&
\NPRsuppDummyIndexOne=\NPRsuppDummyIndexTwo.
\end{cases}
\end{align}
For the complex-conjugate $2\!\times\!2$ blocks below, we will additionally need two elementary trigonometric scalar integrals: for real $\NPRsuppDummyIndexOne$ and $\NPRsuppDummyIndexTwo$,
\begin{align}\label{NPR.eq:elementaryIntegrals}
\begin{aligned}
\NPRsuppCosIntegral(\NPRsuppDummyIndexOne, \NPRsuppDummyIndexTwo) &:= \int_0^1 e^{\NPRsuppDummyIndexOne s}\cos(\NPRsuppDummyIndexTwo s)\,\dd s = \frac{e^{\NPRsuppDummyIndexOne}\bigl[\NPRsuppDummyIndexOne\cos(\NPRsuppDummyIndexTwo) + \NPRsuppDummyIndexTwo\sin(\NPRsuppDummyIndexTwo)\bigr] - \NPRsuppDummyIndexOne}{\NPRsuppDummyIndexOne^2 + \NPRsuppDummyIndexTwo^2}, \\[4pt]
\NPRsuppSinIntegral(\NPRsuppDummyIndexOne, \NPRsuppDummyIndexTwo) &:= \int_0^1 e^{\NPRsuppDummyIndexOne s}\sin(\NPRsuppDummyIndexTwo s)\,\dd s = \frac{e^{\NPRsuppDummyIndexOne}\bigl[\NPRsuppDummyIndexOne\sin(\NPRsuppDummyIndexTwo) - \NPRsuppDummyIndexTwo\cos(\NPRsuppDummyIndexTwo)\bigr] + \NPRsuppDummyIndexTwo}{\NPRsuppDummyIndexOne^2 + \NPRsuppDummyIndexTwo^2},
\end{aligned}
\end{align}
valid whenever $\NPRsuppDummyIndexOne^2 + \NPRsuppDummyIndexTwo^2 > 0$, with $\NPRsuppCosIntegral(0,0) = 1$ and $\NPRsuppSinIntegral(0,0) = 0$ by continuous extension. They satisfy the parity relations
\begin{align}\label{NPR.eq:parity}
\NPRsuppCosIntegral(\NPRsuppDummyIndexOne, -\NPRsuppDummyIndexTwo) = \NPRsuppCosIntegral(\NPRsuppDummyIndexOne, \NPRsuppDummyIndexTwo), \qquad \NPRsuppSinIntegral(\NPRsuppDummyIndexOne, -\NPRsuppDummyIndexTwo) = -\NPRsuppSinIntegral(\NPRsuppDummyIndexOne, \NPRsuppDummyIndexTwo),
\end{align}
and reduce to the scalar divided difference~\eqref{NPR.eq:varphiDef} through $\varphi(\NPRsuppDummyIndexOne, \NPRsuppDummyIndexTwo) = e^{\NPRsuppDummyIndexOne}\,\NPRsuppCosIntegral(\NPRsuppDummyIndexTwo - \NPRsuppDummyIndexOne, 0)$.
The formulas below use $\NPRsuppCosIntegral$ and $\NPRsuppSinIntegral$ as the real-valued scalar building blocks; the scalar divided difference $\varphi$ is kept only for the purely scalar case.

\paragraph{Block notation.}
For $1\!\times\!1$ blocks, we write
\begin{align}
  \NPRBlk{\NPRBlockOneIndex}=\NPRBlockScalarOne \quad \NPRBlk{\NPRBlockTwoIndex}=\NPRBlockScalarTwo
\end{align}
For $2\!\times\!2$ complex-conjugate blocks, with $\mathbf{I}_2$ the identity and
$\mathbf{J}_2 := \bigl(\begin{smallmatrix}0 & 1 \\ -1 & 0\end{smallmatrix}\bigr)$ satisfying $\mathbf{J}_2^2 = -\mathbf{I}_2$, we write
\begin{equation}\label{NPRsupp.complexBlockStructure}
\begin{aligned}
\NPRBlk{\NPRBlockOneIndex}
&=
\begin{pmatrix}
\NPRRealPartOne & \NPRImaginaryPartOne \\
-\NPRImaginaryPartOne & \NPRRealPartOne
\end{pmatrix}
=
\NPRRealPartOne\,\mathbf{I}_2+\NPRImaginaryPartOne\,\mathbf{J}_2
,
\\
\NPRBlk{\NPRBlockTwoIndex}
&=
\begin{pmatrix}
\NPRRealPartTwo & \NPRImaginaryPartTwo \\
-\NPRImaginaryPartTwo & \NPRRealPartTwo
\end{pmatrix}
=
\NPRRealPartTwo\,\mathbf{I}_2+\NPRImaginaryPartTwo\,\mathbf{J}_2
,
\end{aligned}
\end{equation}
so that $\NPRRealPartOne,\NPRRealPartTwo$ denote the real parts and $\NPRImaginaryPartOne,\NPRImaginaryPartTwo$ the corresponding signed imaginary parts in real block form.

\paragraph{Connection with the main manuscript} The generic block integral from Equation~\eqref{NPRsupp.eq:blockDiagonal} is the building block for the adjoint Fr\'echet derivative used in the derivation of the exact reverse-mode gradient in the main manuscript.
Throughout this section we write the closed forms using untransposed blocks $\NPRBlk{\NPRBlockOneIndex}$ and $\NPRBlk{\NPRBlockTwoIndex}$.
In the adjoint application in the main manuscript, the same formulas apply to the transposed blocks by replacing the imaginary part of each $2\times2$ block by its negative; for example, $\NPRImaginaryPartOne\mapsto-\NPRImaginaryPartOne$ and $\NPRImaginaryPartTwo\mapsto-\NPRImaginaryPartTwo$.

From an implementation perspective, the appropriate use of the kernels presented in this section depends on the model setting.
When the focus is on a single CTMC, it is convenient to apply the closed-form kernels directly to the adjoint rank-one contractions; therefore, the matrix $\NPRsuppFrechetDirectionBlock{}$ used in Equation~\eqref{NPRsupp.eq:blockDiagonal} is never materialized.
In the multi-site setting that we discuss briefly in the application section of the main manuscript, site-wise adjoint contributions can be
accumulated into a dense adjoint matrix in the block basis, reducing the number
of required block-pair kernel evaluations.

\subsection{Block pairs involving scalar blocks}\label{npr.sec:oneDimBlockPairs}
\paragraph{$1\times1$ - $1\times1$ blocks} If both blocks are scalar, that is,
$\NPRBlk{\NPRBlockOneIndex}=\NPRBlockScalarOne$
and
$\NPRBlk{\NPRBlockTwoIndex}=\NPRBlockScalarTwo$,
then integral \eqref{NPRsupp.eq:blockDiagonal} evaluates to
\begin{align}\label{NPR.eq:case11Result}
\varphi(\NPRtimeValue\NPRBlockScalarOne,\,\NPRtimeValue\NPRBlockScalarTwo)\,\NPRsuppFrechetDirectionBlock{\NPRBlockOneIndex \NPRBlockTwoIndex},
\end{align}
where $\NPRsuppFrechetDirectionBlock{\NPRBlockOneIndex \NPRBlockTwoIndex}$ is also a scalar.

\paragraph{$1\times1$ - $2\times2$ blocks} If $\NPRBlk{\NPRBlockOneIndex}=\NPRBlockScalarOne$ is scalar and
$\NPRBlk{\NPRBlockTwoIndex} = \NPRRealPartTwo\,\mathbf{I}_2 + \NPRImaginaryPartTwo\,\mathbf{J}_2\in\mathbb{R}^{2\times2}$ is a complex-conjugate block with eigenvalues $\NPRRealPartTwo \pm \mathrm{i}\NPRImaginaryPartTwo$, then writing $\NPRsuppFrechetDirectionBlock{\NPRBlockOneIndex \NPRBlockTwoIndex} = (\NPRsuppFrechetDirectionBlockElement{1}, \NPRsuppFrechetDirectionBlockElement{2})$ as a $1\!\times\!2$ row vector, \eqref{NPRsupp.eq:blockDiagonal} is equal to
\begin{align}\label{NPR.eq:case12Result}
e^{\NPRtimeValue\NPRBlockScalarOne}\,
\NPRsuppFrechetDirectionBlock{\NPRBlockOneIndex \NPRBlockTwoIndex}\,
\begin{pmatrix}
\NPRsuppCosIntegral\bigl(\NPRtimeValue(\NPRRealPartTwo - \NPRBlockScalarOne),\,\NPRtimeValue\NPRImaginaryPartTwo\bigr) & \NPRsuppSinIntegral\bigl(\NPRtimeValue(\NPRRealPartTwo - \NPRBlockScalarOne),\,\NPRtimeValue\NPRImaginaryPartTwo\bigr) \\[0.3ex]
-\NPRsuppSinIntegral\bigl(\NPRtimeValue(\NPRRealPartTwo - \NPRBlockScalarOne),\,\NPRtimeValue\NPRImaginaryPartTwo\bigr) & \NPRsuppCosIntegral\bigl(\NPRtimeValue(\NPRRealPartTwo - \NPRBlockScalarOne),\,\NPRtimeValue\NPRImaginaryPartTwo\bigr)
\end{pmatrix}.
\end{align}
This follows by factoring the scalar exponential and integrating $e^{s\NPRtimeValue(\NPRRealPartTwo-\NPRBlockScalarOne)}\bigl(\cos(s\NPRtimeValue\NPRImaginaryPartTwo)\mathbf{I}_2 + \sin(s\NPRtimeValue\NPRImaginaryPartTwo)\mathbf{J}_2\bigr)$ entrywise.

\paragraph{$2\times2$ - $1\times1$ blocks}  The case $\NPRBlk{\NPRBlockOneIndex}\in\mathbb{R}^{2\times2}$ with eigenvalues $\NPRRealPartOne \pm \mathrm{i}\NPRImaginaryPartOne$, $\NPRBlk{\NPRBlockTwoIndex}=\NPRBlockScalarTwo$ is analogous: writing $\NPRsuppFrechetDirectionBlock{\NPRBlockOneIndex \NPRBlockTwoIndex} = (\NPRsuppFrechetDirectionBlockElement{1}, \NPRsuppFrechetDirectionBlockElement{2})^{\top}$ as a $2\!\times\!1$ column vector and reversing the integration variable in~\eqref{NPRsupp.eq:blockDiagonal} to relocate the scalar exponential to a leading prefactor, the integral equals
\begin{align}\label{NPR.eq:case21Result}
e^{\NPRtimeValue\NPRBlockScalarTwo}\,
\begin{pmatrix}
\NPRsuppCosIntegral\bigl(\NPRtimeValue(\NPRRealPartOne - \NPRBlockScalarTwo),\,\NPRtimeValue\NPRImaginaryPartOne\bigr) & \NPRsuppSinIntegral\bigl(\NPRtimeValue(\NPRRealPartOne - \NPRBlockScalarTwo),\,\NPRtimeValue\NPRImaginaryPartOne\bigr) \\[0.3ex]
-\NPRsuppSinIntegral\bigl(\NPRtimeValue(\NPRRealPartOne - \NPRBlockScalarTwo),\,\NPRtimeValue\NPRImaginaryPartOne\bigr) & \NPRsuppCosIntegral\bigl(\NPRtimeValue(\NPRRealPartOne - \NPRBlockScalarTwo),\,\NPRtimeValue\NPRImaginaryPartOne\bigr)
\end{pmatrix}
\NPRsuppFrechetDirectionBlock{\NPRBlockOneIndex \NPRBlockTwoIndex}.
\end{align}

\subsection{\texorpdfstring{Complex-conjugate \(2\times2\) block pairs}{Complex-conjugate 2x2 block pairs}}\label{npr.sec:complexConjBlockPairs}
The most challenging case arises when both blocks have the $2\!\times\!2$ complex-conjugate form introduced in Equation~\eqref{NPRsupp.complexBlockStructure}, with eigenvalues $\NPRRealPartOne \pm \mathrm{i}\NPRImaginaryPartOne$ ($\NPRImaginaryPartOne \neq 0 $) and $\NPRRealPartTwo \pm \mathrm{i}\NPRImaginaryPartTwo$ ($\NPRImaginaryPartTwo \neq 0)$.
The matrix exponential of $\NPRBlk{\NPRBlockOneIndex}$ scaled by $\NPRtimeValue$ can be computed using the following trigonometric representation (using the $\mathbf{I}_2,\mathbf{J}_2$ representation in~\eqref{NPRsupp.complexBlockStructure}):
\begin{align}\label{NPR.eq:supp-exp-block}
e^{s\NPRtimeValue\NPRBlk{\NPRBlockOneIndex}}
= e^{\NPRtimeValue s\NPRRealPartOne}
\begin{pmatrix}
\cos(\NPRtimeValue s\NPRImaginaryPartOne) & \sin(\NPRtimeValue s\NPRImaginaryPartOne) \\
-\sin(\NPRtimeValue s\NPRImaginaryPartOne) & \cos(\NPRtimeValue s\NPRImaginaryPartOne)
\end{pmatrix}.
\end{align}
A similar expression holds for $\NPRBlk{\NPRBlockTwoIndex}$.
\paragraph{Complex coordinate decomposition.}
Write the $(\NPRBlockOneIndex,\NPRBlockTwoIndex)$-submatrix of the direction as
\begin{align}
\NPRsuppFrechetDirectionBlock{\NPRBlockOneIndex \NPRBlockTwoIndex} =
\begin{pmatrix}
\NPRsuppFrechetDirectionBlockElement{11} & \NPRsuppFrechetDirectionBlockElement{12} \\
\NPRsuppFrechetDirectionBlockElement{21} & \NPRsuppFrechetDirectionBlockElement{22}
\end{pmatrix}.
\end{align}
Define
\begin{align}
z_- &= \frac{\NPRsuppFrechetDirectionBlockElement{11}+\NPRsuppFrechetDirectionBlockElement{22}}{2} + \mathrm{i}\,\frac{\NPRsuppFrechetDirectionBlockElement{12}-\NPRsuppFrechetDirectionBlockElement{21}}{2},
\\[4pt]
z_+ &= \frac{\NPRsuppFrechetDirectionBlockElement{11}-\NPRsuppFrechetDirectionBlockElement{22}}{2} + \mathrm{i}\,\frac{\NPRsuppFrechetDirectionBlockElement{12}+\NPRsuppFrechetDirectionBlockElement{21}}{2}.
\end{align}
These two complex coordinates span invariant one-dimensional subspaces of the
block kernel.
For each $s\in[0,1]$, the joint action of $e^{(1-s)\NPRtimeValue\NPRBlk{\NPRBlockOneIndex}}$ on the left and $e^{s\NPRtimeValue\NPRBlk{\NPRBlockTwoIndex}}$ on the right multiplies $z_-$ by
$e^{(1-s)\NPRtimeValue(\NPRRealPartOne+\mathrm{i}\NPRImaginaryPartOne)}\,e^{s\NPRtimeValue(\NPRRealPartTwo+\mathrm{i}\NPRImaginaryPartTwo)}$
and $z_+$ by
$e^{(1-s)\NPRtimeValue(\NPRRealPartOne-\mathrm{i}\NPRImaginaryPartOne)}\,e^{s\NPRtimeValue(\NPRRealPartTwo+\mathrm{i}\NPRImaginaryPartTwo)}$.
Integrating over $s\in[0,1]$ yields
\begin{align}
z_-^{\mathrm{out}} &= e^{\NPRtimeValue(\NPRRealPartOne+\mathrm{i}\NPRImaginaryPartOne)}\,
\NPRsuppComplexIntegral\!\bigl(\NPRtimeValue(\NPRRealPartTwo-\NPRRealPartOne),\,\NPRtimeValue(\NPRImaginaryPartTwo-\NPRImaginaryPartOne)\bigr) \,z_-,
\\
z_+^{\mathrm{out}} &= e^{\NPRtimeValue(\NPRRealPartOne-\mathrm{i}\NPRImaginaryPartOne)}\,
\NPRsuppComplexIntegral\!\bigl(\NPRtimeValue(\NPRRealPartTwo-\NPRRealPartOne),\,\NPRtimeValue(\NPRImaginaryPartOne+\NPRImaginaryPartTwo)\bigr)\,z_+,
\end{align}
where the temporary complex shorthand is
\begin{align}
\NPRsuppComplexIntegral(\NPRsuppDummyIndexOne,\NPRsuppDummyIndexTwo)
&:=
\int_0^1 e^{(\NPRsuppDummyIndexOne+\mathrm{i}\NPRsuppDummyIndexTwo)s}\,\dd s
 =
\NPRsuppCosIntegral(\NPRsuppDummyIndexOne,\NPRsuppDummyIndexTwo)+\mathrm{i}\,\NPRsuppSinIntegral(\NPRsuppDummyIndexOne,\NPRsuppDummyIndexTwo)
\label{NPR.eq:complex-J}
\end{align}
(equivalently, $\NPRsuppComplexIntegral(\NPRsuppDummyIndexOne,\NPRsuppDummyIndexTwo)=(e^{\NPRsuppDummyIndexOne+\mathrm{i}\NPRsuppDummyIndexTwo}-1)/(\NPRsuppDummyIndexOne+\mathrm{i}\NPRsuppDummyIndexTwo)$ when $\NPRsuppDummyIndexOne+\mathrm{i}\NPRsuppDummyIndexTwo\neq0$).
Equation~\eqref{NPR.eq:complex-J} shows that every
\(2\times2\)--\(2\times2\) block-pair kernel reduces, up to the fixed
coordinate change above, to two complex scalar multiplications, avoiding
any generic \(4\times4\) Padé-based matrix exponential or dense Sylvester solve.

\paragraph{Explicit real-valued closed form.}
For implementation in real arithmetic, it is convenient to expand the complex-coordinate result above into an explicit real-valued formula in terms of the elementary integrals~\eqref{NPR.eq:elementaryIntegrals}.
Factoring $e^{(1-s)\NPRtimeValue\NPRBlk{\NPRBlockOneIndex}} = e^{\NPRtimeValue\NPRBlk{\NPRBlockOneIndex}}\,e^{-s\NPRtimeValue\NPRBlk{\NPRBlockOneIndex}}$ in~\eqref{NPRsupp.eq:blockDiagonal} pulls the constant matrix prefactor $e^{\NPRtimeValue\NPRBlk{\NPRBlockOneIndex}} = e^{\NPRtimeValue\NPRRealPartOne}\bigl(\begin{smallmatrix}\cos(\NPRtimeValue\NPRImaginaryPartOne) & \sin(\NPRtimeValue\NPRImaginaryPartOne) \\ -\sin(\NPRtimeValue\NPRImaginaryPartOne) & \cos(\NPRtimeValue\NPRImaginaryPartOne)\end{smallmatrix}\bigr)$ outside the integral, yielding the two-stage form
\begin{align}\label{NPR.eq:case22TwoStage}
\int_0^1 e^{(1-s)\NPRtimeValue\NPRBlk{\NPRBlockOneIndex}}\NPRsuppFrechetDirectionBlock{\NPRBlockOneIndex \NPRBlockTwoIndex}\,e^{s\NPRtimeValue\NPRBlk{\NPRBlockTwoIndex}}\,\dd s
= e^{\NPRtimeValue\NPRRealPartOne}
\begin{pmatrix}
\cos(\NPRtimeValue\NPRImaginaryPartOne) & \sin(\NPRtimeValue\NPRImaginaryPartOne) \\
-\sin(\NPRtimeValue\NPRImaginaryPartOne) & \cos(\NPRtimeValue\NPRImaginaryPartOne)
\end{pmatrix}\,
\mathbf{Z}_{\NPRBlockOneIndex \NPRBlockTwoIndex},
\end{align}
where the rotation-free $\mathbf{Z}_{\NPRBlockOneIndex \NPRBlockTwoIndex}$ collects the $s$-dependent part of the integrand and is obtained by multiplying out the triple product and applying the product-to-sum identities to reduce $\cos(s\NPRtimeValue\NPRImaginaryPartOne)\cos(s\NPRtimeValue\NPRImaginaryPartTwo)$, $\sin(s\NPRtimeValue\NPRImaginaryPartOne)\sin(s\NPRtimeValue\NPRImaginaryPartTwo)$, $\sin(s\NPRtimeValue\NPRImaginaryPartOne)\cos(s\NPRtimeValue\NPRImaginaryPartTwo)$, $\cos(s\NPRtimeValue\NPRImaginaryPartOne)\sin(s\NPRtimeValue\NPRImaginaryPartTwo)$ to combinations of $\cos(s\NPRtimeValue(\NPRImaginaryPartOne\pm\NPRImaginaryPartTwo))$ and $\sin(s\NPRtimeValue(\NPRImaginaryPartOne\pm\NPRImaginaryPartTwo))$. Abbreviating
\begin{align}\label{NPR.eq:case22ICIS}
\NPRsuppCosIntegral^{\pm} := \NPRsuppCosIntegral\bigl(\NPRtimeValue(\NPRRealPartTwo - \NPRRealPartOne),\,\NPRtimeValue(\NPRImaginaryPartOne \pm \NPRImaginaryPartTwo)\bigr),
\qquad
\NPRsuppSinIntegral^{\pm} := \NPRsuppSinIntegral\bigl(\NPRtimeValue(\NPRRealPartTwo - \NPRRealPartOne),\,\NPRtimeValue(\NPRImaginaryPartOne \pm \NPRImaginaryPartTwo)\bigr),
\end{align}
the four entries of $\mathbf{Z}_{\NPRBlockOneIndex \NPRBlockTwoIndex}$ are
\begin{align}\label{NPR.eq:case22ZEntries}
\begin{aligned}
\bigl[\mathbf{Z}_{\NPRBlockOneIndex \NPRBlockTwoIndex}\bigr]_{1,1}
&= \tfrac{1}{2}\bigl[\,(\NPRsuppFrechetDirectionBlockElement{11}+\NPRsuppFrechetDirectionBlockElement{22})\,\NPRsuppCosIntegral^- + (\NPRsuppFrechetDirectionBlockElement{11}-\NPRsuppFrechetDirectionBlockElement{22})\,\NPRsuppCosIntegral^+ + (\NPRsuppFrechetDirectionBlockElement{12}-\NPRsuppFrechetDirectionBlockElement{21})\,\NPRsuppSinIntegral^- - (\NPRsuppFrechetDirectionBlockElement{12}+\NPRsuppFrechetDirectionBlockElement{21})\,\NPRsuppSinIntegral^+\,\bigr], \\
\bigl[\mathbf{Z}_{\NPRBlockOneIndex \NPRBlockTwoIndex}\bigr]_{1,2}
&= \tfrac{1}{2}\bigl[\,(\NPRsuppFrechetDirectionBlockElement{12}-\NPRsuppFrechetDirectionBlockElement{21})\,\NPRsuppCosIntegral^- + (\NPRsuppFrechetDirectionBlockElement{12}+\NPRsuppFrechetDirectionBlockElement{21})\,\NPRsuppCosIntegral^+ - (\NPRsuppFrechetDirectionBlockElement{11}+\NPRsuppFrechetDirectionBlockElement{22})\,\NPRsuppSinIntegral^- + (\NPRsuppFrechetDirectionBlockElement{11}-\NPRsuppFrechetDirectionBlockElement{22})\,\NPRsuppSinIntegral^+\,\bigr], \\
\bigl[\mathbf{Z}_{\NPRBlockOneIndex \NPRBlockTwoIndex}\bigr]_{2,1}
&= \tfrac{1}{2}\bigl[\,(\NPRsuppFrechetDirectionBlockElement{21}-\NPRsuppFrechetDirectionBlockElement{12})\,\NPRsuppCosIntegral^- + (\NPRsuppFrechetDirectionBlockElement{21}+\NPRsuppFrechetDirectionBlockElement{12})\,\NPRsuppCosIntegral^+ + (\NPRsuppFrechetDirectionBlockElement{11}+\NPRsuppFrechetDirectionBlockElement{22})\,\NPRsuppSinIntegral^- + (\NPRsuppFrechetDirectionBlockElement{11}-\NPRsuppFrechetDirectionBlockElement{22})\,\NPRsuppSinIntegral^+\,\bigr], \\
\bigl[\mathbf{Z}_{\NPRBlockOneIndex \NPRBlockTwoIndex}\bigr]_{2,2}
&= \tfrac{1}{2}\bigl[\,(\NPRsuppFrechetDirectionBlockElement{11}+\NPRsuppFrechetDirectionBlockElement{22})\,\NPRsuppCosIntegral^- + (\NPRsuppFrechetDirectionBlockElement{22}-\NPRsuppFrechetDirectionBlockElement{11})\,\NPRsuppCosIntegral^+ + (\NPRsuppFrechetDirectionBlockElement{12}-\NPRsuppFrechetDirectionBlockElement{21})\,\NPRsuppSinIntegral^- + (\NPRsuppFrechetDirectionBlockElement{12}+\NPRsuppFrechetDirectionBlockElement{21})\,\NPRsuppSinIntegral^+\,\bigr].
\end{aligned}
\end{align}
The two formulations are equivalent: the complex multipliers are
\begin{align}\label{NPR.eq:complexRealEquivalence}
m_- = e^{\NPRtimeValue(\NPRRealPartOne+\mathrm{i}\NPRImaginaryPartOne)}\,(\NPRsuppCosIntegral^{-} - \mathrm{i}\,\NPRsuppSinIntegral^{-}),
\qquad
m_+ = e^{\NPRtimeValue(\NPRRealPartOne-\mathrm{i}\NPRImaginaryPartOne)}\,(\NPRsuppCosIntegral^{+} + \mathrm{i}\,\NPRsuppSinIntegral^{+}),
\end{align}
the asymmetric sign on $\NPRsuppSinIntegral^{-}$ following from parity~\eqref{NPR.eq:parity}, since $\NPRsuppComplexIntegral\!\bigl(\NPRtimeValue(\NPRRealPartTwo-\NPRRealPartOne),\NPRtimeValue(\NPRImaginaryPartTwo-\NPRImaginaryPartOne)\bigr)=\NPRsuppCosIntegral^- - \mathrm{i}\NPRsuppSinIntegral^-$ under our real-integral convention $\NPRsuppDummyIndexTwo = \NPRtimeValue(\NPRImaginaryPartOne \pm \NPRImaginaryPartTwo)$ for the ``$\pm$'' frequencies.
Substituting~\eqref{NPR.eq:complexRealEquivalence} into the reconstruction formula above recovers~\eqref{NPR.eq:case22ZEntries}.
The real form uses the same building blocks across all four block-pair cases of Sections~\ref{npr.sec:oneDimBlockPairs}--\ref{npr.sec:complexConjBlockPairs} and is more amenable to vectorized or GPU-targeted execution.

\subsection{Stable evaluation of the scalar integrals}

The main numerical special case behind~\eqref{NPR.eq:complex-J} is a near-zero
denominator $\NPRsuppDummyIndexOne+\mathrm{i}\NPRsuppDummyIndexTwo\approx 0$.
Direct evaluation of $(e^{\NPRsuppDummyIndexOne+\mathrm{i}\NPRsuppDummyIndexTwo}-1)/(\NPRsuppDummyIndexOne+\mathrm{i}\NPRsuppDummyIndexTwo)$
then suffers from catastrophic cancellation.
The two relevant arguments are
\begin{align}
(\NPRsuppDummyIndexOne_{-},\NPRsuppDummyIndexTwo_{-})
=\bigl(\NPRtimeValue(\NPRRealPartTwo-\NPRRealPartOne),\,\NPRtimeValue(\NPRImaginaryPartTwo-\NPRImaginaryPartOne)\bigr),
\qquad
(\NPRsuppDummyIndexOne_{+},\NPRsuppDummyIndexTwo_{+})
=\bigl(\NPRtimeValue(\NPRRealPartTwo-\NPRRealPartOne),\,\NPRtimeValue(\NPRImaginaryPartOne+\NPRImaginaryPartTwo)\bigr).
\end{align}
The near-zero case occurs when the interacting blocks have nearly equal
eigenvalues ($\NPRRealPartOne\approx \NPRRealPartTwo$, $\NPRImaginaryPartTwo\approx\pm\NPRImaginaryPartOne$).
In this regime the limiting value $\NPRsuppComplexIntegral\to1$, equivalently $\NPRsuppCosIntegral\to1$ and $\NPRsuppSinIntegral\to0$, is used, consistent with
L'H\^opital's rule and the convergent Taylor expansion
\begin{align}
\NPRsuppComplexIntegral(\NPRsuppDummyIndexOne,\NPRsuppDummyIndexTwo)
=
\sum_{n=0}^{\infty}
\frac{(\NPRsuppDummyIndexOne+\mathrm{i}\NPRsuppDummyIndexTwo)^n}{(n+1)!}.
\end{align}
This yields numerically stable evaluation of the Fr\'echet coefficients.

\section{Additional simulation-study diagnostics using exact and approximate gradients} \label{NPRsupp.sec:simulationStudyFollowup}

The simulation study in the main text was designed to evaluate whether the proposed GP-prior CTMC model can recover a nonlinear relationship between a pairwise covariate and the normalized CTMC log-rates.
Here we provide additional inferential diagnostics for that experiment by comparing two implementations of the likelihood gradient: the exact gradient and the approximate gradient introduced in the main text.
Both chains were run for $48$ hours on a single CPU thread on UCLA's Hoffman2 high-performance computing cluster.
Samples were recorded every $100$ iterations, and the first $10\%$ of each chain was discarded as burn-in.
Mixing was assessed using effective sample sizes (ESS), computed with the spectral-density-at-zero estimator of \citet{heidelberger_spectral_1981}.

The two gradient implementations have different computational costs but, as shown in Table~\ref{NPRsupp.tab:simDiagnostics}, lead to nearly identical posterior inference.
In $48$ hours, the exact-gradient chain completed approximately $2{,}270{,}000$ iterations and produced $\approx 20{,}400$ post-burn-in samples.
Over the same wall-clock time, the approximate-gradient chain completed more iterations and produced $\approx 23{,}600$ post-burn-in samples, consistent with the lower per-iteration cost of the approximation reported in the benchmark results.
Both chains mixed well across the monitored parameters.
Across the $272$ off-diagonal rate-matrix entries and the two scalar hyperparameters, ESS values ranged from $520$ to $9{,}760$ under the exact gradient, with median ESS $650$.
Under the approximate gradient, the corresponding ESS values ranged from $670$ to $15{,}400$, with median ESS $800$.
Posterior summaries for the GP length-scale hyperparameter were essentially unchanged across the two implementations: both chains had posterior mean $2.44$, nearly identical medians ($2.34$ versus $2.33$), identical posterior standard deviations ($0.39$), and fully overlapping $95\%$ credible intervals.

The performances of the two gradient approaches are consistent with the structure of the simulated dataset.
The simulated tree has short branch lengths, which is precisely the regime in which the gradient approximation is expected to be most accurate.
Consequently, the approximate gradient produces posterior inference that is effectively indistinguishable from that obtained with the exact gradient, while reducing the computational cost of the analysis (or, alternatively, yielding higher ESS values).

\begin{table}[ht]
\centering
\caption{MCMC diagnostics for the simulation study under the exact and approximate gradients}
\label{NPRsupp.tab:simDiagnostics}
\begin{tabular}{lcc}
\hline
& Exact & Approximate \\
\hline
Post-burn-in samples & $20{,}416$ & $23{,}559$ \\
Minimum ESS          & $520$      & $670$      \\
Median ESS           & $650$      & $800$      \\
\hline
\multicolumn{3}{l}{\textit{GP length-scale hyperparameter}} \\
\hline
Mean                 & $2.44$     & $2.44$     \\
Median               & $2.34$     & $2.33$     \\
Standard deviation   & $0.39$     & $0.39$     \\
$95\%$ credible interval & $[2.01,\, 3.44]$ & $[2.01,\, 3.40]$ \\
\hline
\end{tabular}
\end{table}

\bibliographystyle{apalike}
\bibliography{references}